\documentclass[sigconf, nonacm, authorversion=true]{acmart}

\newcommand\vldbyear{2026}
\newcommand\vldbworkshop{The 2nd Workshop on Vector Databases}
\newcommand\vldbauthors{\authors}
\newcommand\vldbtitle{\shorttitle} 
\newcommand\vldbavailabilityurl{URL_TO_YOUR_ARTIFACTS}
\newcommand\vldbpagestyle{plain} 

\AtBeginDocument{%
  \providecommand\BibTeX{{%
    \normalfont B\kern-0.5em{\scshape i\kern-0.25em b}\kern-0.8em\TeX}}}

\usepackage{listings}
\usepackage{balance}
\usepackage{color}
\usepackage{xcolor}
\usepackage{pifont}

\usepackage{colortbl}
\usepackage{subcaption}

\usepackage{lipsum}
\usepackage{verbatim}
\usepackage{colortbl}
\usepackage{multirow}
\usepackage{soul}
\usepackage{framed}

\usepackage{url}
\usepackage{hyperref}

\usepackage[linesnumbered,ruled,vlined]{algorithm2e}

\usepackage{enumitem}

\newlist{questions}{enumerate}{2}
\setlist[questions,1]{label=\textbf{(Q\arabic*)},ref=\textbf{(Q\arabic*)}}
\setlist[questions,2]{label=(\alph*),ref=\thequestionsi(\alph*)}

\newcommand{\mathcolorbox}[2]{\colorbox{#1}{$\displaystyle #2$}}
\NewDocumentCommand{\codeword}{v}{%
    \texttt{\textcolor{violet}{#1}}%
}

\NewDocumentCommand{\colorcodeword}{vv}{%
    \textbf{\texttt{\textcolor{#1}{#2}}}%
}

\usepackage{xcolor}
\usepackage{framed}

\definecolor{lightgraybox}{gray}{0.95}
\definecolor{bordergray}{gray}{0.75}

\makeatletter
\def\NAT@def@citea{\def\@citea{\NAT@separator}}
\makeatother

\definecolor{lightgray}{gray}{0.95}
\definecolor{lightpeach}{HTML}{FCE5CD}
\definecolor{softpink}{RGB}{245, 220, 225}
\definecolor{softblue}{HTML}{dae8fc}
\definecolor{lightgreen}{HTML}{eafbe5}

\begin{document}



\title[Stop Indexing at Full Precision: Revisiting Clustering for Vector Embeddings]{Stop Indexing at Full Precision:\\ Revisiting Clustering for Vector Embeddings}


\settopmatter{authorsperrow=3}

\author{Leonardo Kuffo}
\affiliation{%
  \institution{CWI}
  \city{Amsterdam}
  \country{The Netherlands}}
\email{lxkr@cwi.nl}

\author{Peter Boncz}
\affiliation{%
  \institution{CWI}
  \city{Amsterdam}
  \country{The Netherlands}}
\email{boncz@cwi.nl}



\begin{abstract}

In this study, we revisit three widely used techniques in vector search and utilize them to optimize vector embedding indexing through clustering: dimensionality reduction, quantization, and dimension pruning. We propose an indexing pipeline in which these techniques are applied \textit{before} clustering, and we focus on how they affect storage footprint, clustering time, and the quality of the resulting centroids for vector search tasks. Our results reveal that using full-precision vectors for clustering is excessive, as even 1-bit codes can achieve near-optimal clustering quality (within 1\% of ideal) while reducing storage requirements by 60x and delivering attractive performance gains (Figure 1). We open-source our implementations at~\href{https://github.com/cwida/SuperKMeans}{https://github.com/cwida/SuperKMeans}.


\end{abstract}

\maketitle

\pagestyle{\vldbpagestyle}
\begingroup\small\noindent\raggedright\textbf{VLDB Workshop Reference Format:}\\
\vldbauthors. \vldbtitle. VLDB \vldbyear\ Workshop: \vldbworkshop.\\ 
\endgroup
\begingroup
\renewcommand\thefootnote{}\footnote{\noindent
This work is licensed under the Creative Commons BY-NC-ND 4.0 International License. Visit \url{https://creativecommons.org/licenses/by-nc-nd/4.0/} to view a copy of this license. For any use beyond those covered by this license, obtain permission by emailing \href{mailto:info@vldb.org}{info@vldb.org}. Copyright is held by the owner/author(s). Publication rights licensed to the VLDB Endowment. \\
\raggedright Proceedings of the VLDB Endowment. 
ISSN 2150-8097. \\
}\addtocounter{footnote}{-1}\endgroup

\section{Introduction}\label{sec:intro}

Nowadays, clustering algorithms (e.g., $k$-means~\cite{kmeans}) are utilized in \textit{approximate vector similarity search} (VSS) to index large collections of high-dimensional vectors~\cite{quake, scannwhitepaper, lorann, anisotropic, rabitqext, turbopuffer, faisspaper, rabitqgpu, micronn, superkmeans, databricks2026decoupled, panorama, plaid}. VSS consists of finding the vectors in a collection that are the most similar to a given query vector, based on a distance or similarity metric. When a vector collection is indexed using clustering, the resulting centroids act as entry points to guide queries to the clusters that are most likely to contain their nearest neighbors. Thus, reducing search latency by avoiding accessing the majority of the collection.
This indexing method remains a popular option across vector systems due to its scalability and advantages in data management tasks compared to other index families, such as graph-based ones~\cite{micronn, quake, zhu2025experimental, spfresh, turbopuffer, scannwhitepaper, crackivf, cloudvss}.


\begin{figure}[t!]
\centering
\includegraphics[width=1.0\columnwidth]{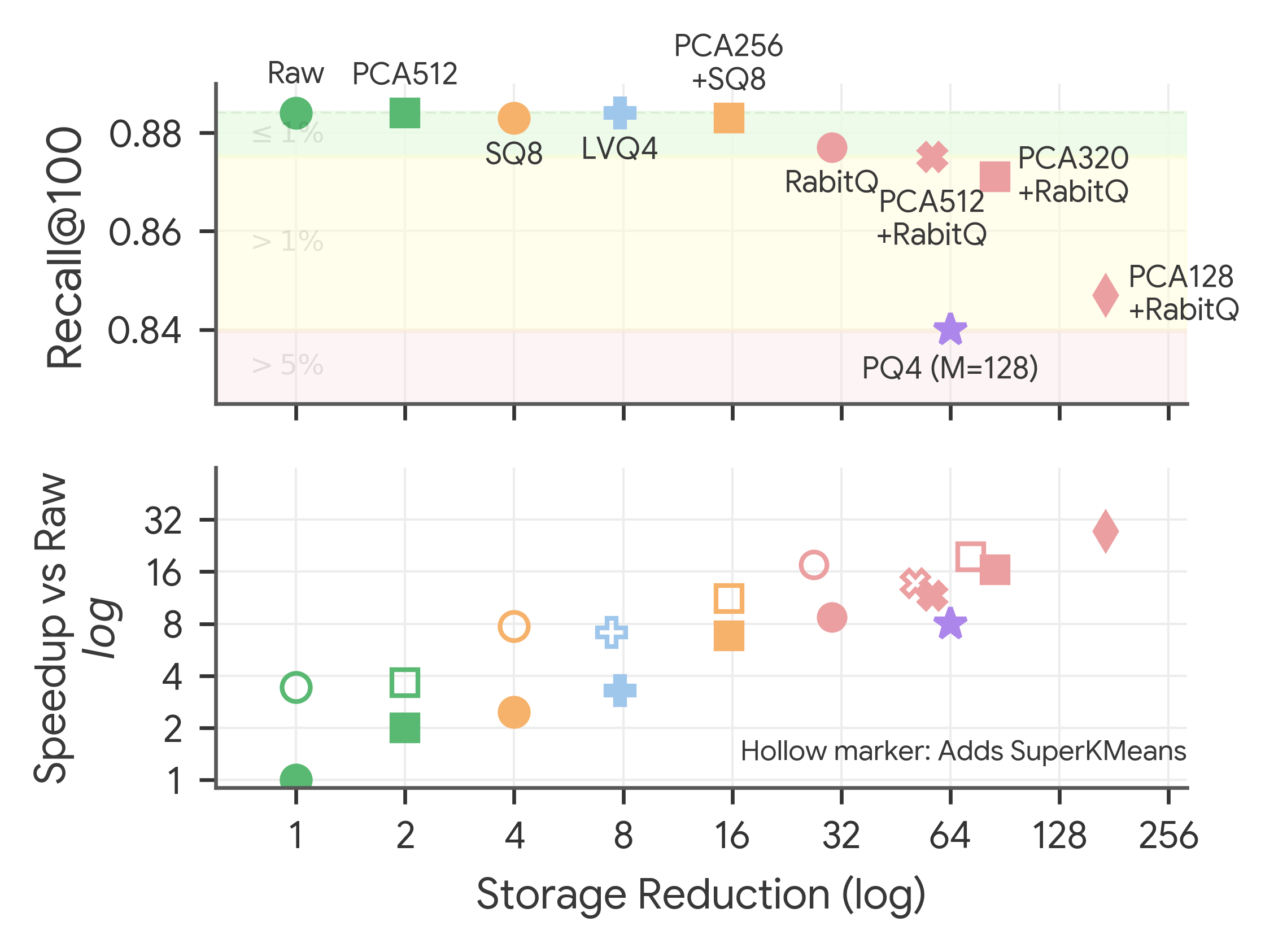}
\vspace*{-9mm}
\caption{Clustering performance for all competitors in 10M 1024-dimensional Cohere embeddings. Clustering is resilient to approximation techniques, achieving near-optimal clustering quality with 60x storage reduction (top). Clustering speed also improves, even more with SuperKMeans (shown with hollow markers), which brings attractive speedups without hurting index quality (bottom). \colorbox[HTML]{eafbe5}{Green}, \colorbox[HTML]{fffddb}{yellow}, and \colorbox[HTML]{f9e0e0}{red} indicate $\leq$1\%, $>$1\%, and $>$5\% away, resp., from the results of clustering with raw (full-precision) vectors. The evaluation framework is presented in Section~\ref{sec:eval}.
}
\label{fig:opening}
\vspace*{-4mm}
\end{figure}


Clustering poses major challenges compared to VSS. First, it is not only memory-bound but also heavily compute-bound~\cite{kuffo2025bang, superkmeans, matsui2017pqk, martinico2026efficient}. Furthermore, it requires accessing the entire collection of raw vectors multiple times, whereas VSS prunes most of the vector collection on every query. As a result, clustering remains a critical step that bottlenecks user queries until the index is built. 
Despite this, most research efforts focus on search algorithms, with few addressing vector indexing via clustering~\cite{matsui2017pqk, superkmeans, chen2025vectorchord100m, kmeanselastic, wang2026lindormvector, martinico2026efficient}.  


In this work, we revisit three widely used techniques for vector search and use them for vector indexing via clustering: \textbf{dimensionality reduction} (JLT~\cite{jltkmeans, jltlemma, chen2025vectorchord100m}, PCA~\cite{pca, leanvec, gleanvec, happymarriage, wang2026lindormvector}, Matryoshka vectors~\cite{kusupati2022matryoshka}), \textbf{quantization} (SQ~\cite{faisspaper}, LVQ~\cite{lvq}, RabitQ~\cite{rabitq, rabitqext, rabitqgpu}, PQ~\cite{pqkmeans_code, pqivf}), and \textbf{dimension pruning}~\cite{pdx, dcosurvey1, dcosurvey2, adsampling, dade, panorama}. We propose an indexing pipeline that applies these techniques \textit{before} clustering, resulting in attractive performance gains and storage reductions without sacrificing index quality. Our pipeline design \textit{differs} from most systems, where clustering uses full-precision vectors and quantization occurs only after clustering~\cite{lance, milvus, faisspaper, rabitqgpu}. 


Our results show that using full-precision vectors for clustering is excessive, as even using 1-bit RabitQ codes of PCA-projected vectors achieves near-optimal clustering quality (less than 1\% degradation) with 60x storage reduction. As part of our contributions, we adapt dimension-pruning techniques (SuperKMeans~\cite{superkmeans}, ADSampling~\cite{adsampling}) to integrate with quantization schemes to further accelerate clustering without sacrificing quality. Finally, we explore several design decisions across the indexing pipeline that affect performance and the quality of clusters for vector search tasks. 



\section{Preliminaries}\label{sec:preliminaries}



\subsection{Vector Search}
Nearest Neighbor Search (NNS) consists of finding the closest vectors in a collection to a given query vector, based on a distance or similarity metric. Finding exact answers is often impractical due to the large compute and storage requirements. However, modern applications such as RAG and Recommender Systems powered by NNS typically do not require exact answers as approximate answers are ``good enough". This is known as Approximate Nearest Neighbor Search (ANNS), or Vector Similarity Search (VSS). The approximate nature of VSS enables the use of techniques that reduce storage and compute requirements during query time, such as quantization~\cite{faisspaper, lvq, pqivf, rabitq, rabitqext, zandieh2025turboquant}, dimension pruning~\cite{dcosurvey1, dcosurvey2, adsampling, suco}, and dimensionality reduction~\cite{kusupati2022matryoshka, jltkmeans, pca, happymarriage, li2025saq}.  

These techniques are usually combined with \textit{approximate indexes}~\cite{gou2025symphonyqg, pqivf, rabitqext, lorann} that guide queries toward the most promising vectors in the collection. Among vector indexes, two main families exist: partition-based~\cite{lorann, scannwhitepaper, soar, quake, wei2026pdet, pqivf, spann} and graph-based~\cite{elpis, aisaq, diskann, hnsw, gou2025symphonyqg}. 
While graph-based indexes can answer queries with fewer distance comparisons than partition-based indexes~\cite{distancecomputationshigheringraphs}, they take a considerably higher amount of time to build~\cite{elpis, superkmeans, wei2026pdet, crackivf}. Furthermore, they introduce additional complexity in data management tasks, such as ingestion and updates~\cite{diskannupdates, spfresh, spann}, as well as when combining VSS with traditional SQL-like filtering~\cite{patel2024acorn, gollapudi2023filtered, lu2026depth}. This makes partition-based indexes an attractive option for large collections and distributed or cloud environments~\cite{turbopuffer, bann, cloudvss, scannwhitepaper, quake, databricks2026decoupled}.

\subsection{The Role of Clustering in Vector Search}

\begin{figure}[t!]
\centering
\includegraphics[width=1.0\columnwidth]{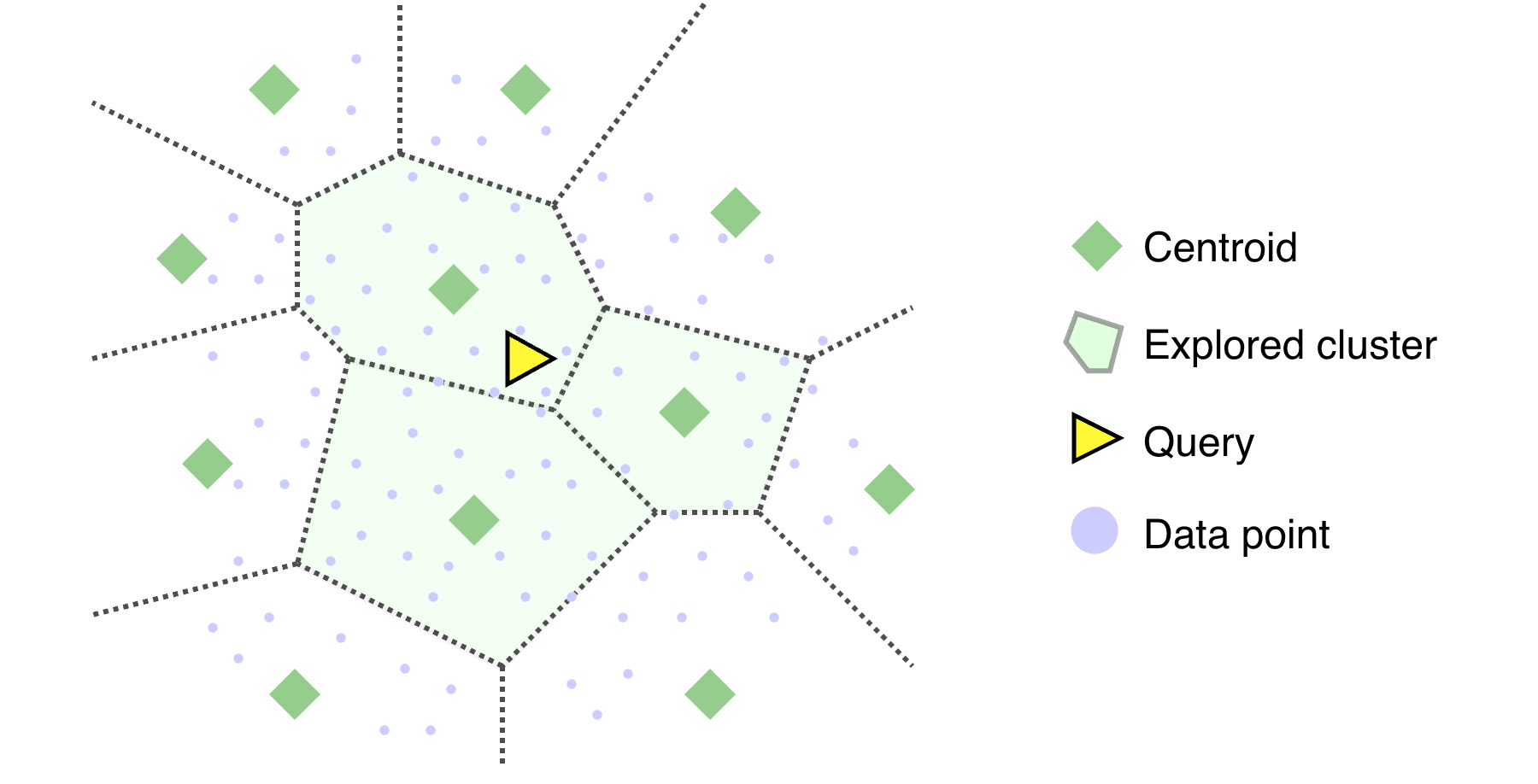}
\vspace*{-3.0mm}
\caption{Example of an IVF index search where only the points in the clusters closest to the query are explored. The clusters are defined via k-means.}
\vspace*{-1.0mm}
\label{fig:ivf}
\end{figure}

Partition-based indexes, such as the widely used Inverted-Files (IVF), organize a collection of vectors into clusters through clustering methods like $k$-means~\cite{faisspaper, lorann, scannwhitepaper}. During query time, the distance metric is first evaluated between the query $q$ and the centroids $Y$. Then, the vectors within the nearest clusters are chosen for evaluation (Figure ~\ref{fig:ivf}). By adjusting the number of explored clusters, one can balance search speed with quality~\cite{faisspaper, milvus}. Typically, the number of centroids is set around $\sqrt{N}$~\cite{faisspaper, milvus}, resulting in higher $k$ values compared to other use cases of clustering~\cite{marigold, kmeansplusplus, zhakubayev2024using}. 

Despite clustering being orders of magnitude faster than graph-index construction, it remains a crucial bottleneck, as users experience degraded performance or are unable to query the collection until an index is built~\cite{crackivf}. Additionally, the clustering of vector embeddings introduces several challenges compared to VSS. First, it requires multiple accesses to the entire vector collection (or a large subsample~\cite{superkmeans, chen2025vectorchord100m}), whereas vector search typically accesses only portions of a collection with each query. Second, unlike vector search, which is mostly data-access bound~\cite{kuffo2025bang}, clustering is also heavily compute-bound due to the higher number of computations needed. On top of that, the increasing demand for vectors as first-class citizens in database systems has created new scenarios in which low-latency, low-memory indexing is essential. For instance, in a situation where an attribute-based filter is applied to a vector column~\cite{lu2026depth}, with the resulting vectors then utilized to perform a similarity \codeword{JOIN}~\cite{simjoin} with a different vector column from another table, the ability to create an index on the fly could enhance the performance of the similarity \codeword{JOIN} operator if it can be created quickly enough.

\subsection{Clustering in Vector Systems}\label{sec:prelim:systems}

The implementation of clustering found in vector systems (e.g., cuVS~\cite{cuvs}, FAISS~\cite{faisspaper}, Milvus~\cite{milvus},  VectorChord~\cite{chen2025vectorchord100m}, Vortex~\cite{vortex}, LanceDB~\cite{lance}, Elastic~\cite{kmeanselastic}) follows a Lloyd $k$-means algorithm~\cite{kmeans} or a hierarchical variant of it. In brief, $k$-means follows these steps: 

\vspace{3mm}\noindent \textbf{STEP 1. Centroids Initialization: } Centroids are initialized by randomly sampling $k$ vectors from the collection~\cite{forgy1965cluster}. More sophisticated initializations, such as $k$-means++~\cite{kmeansplusplus}, have proven ineffective for vector embedding datasets due to their marginal improvements while incurring substantially higher runtime~\cite{superkmeans}. 

\vspace{3mm}\noindent \textbf{STEP 2. Determining Assignments: } Assignments are determined using a \textit{distance metric} that identifies which point $x$ in the collection $X$ is closest to which centroid $y$ of $Y$. The L2 Euclidean distance is the most commonly used. This step is the main bottleneck of clustering, as it requires computing pairwise distances between every $x$ in $X$ and every $y$ in $Y$. The most efficient way to do this is through a General Matrix Multiplication (GEMM) of the data points and centroids ($\mathcolorbox{lightpeach}{X \boldsymbol{\cdot} Y}$). GEMM routines are highly efficient due to their optimized data access patterns, vectorization (SIMD), efficient cache usage, and multi-threading. 

\vspace{3mm}\noindent\textbf{STEP 3. Updating Centroids: } Centroids are updated by averaging every vector $x$ assigned to them. When centroids have zero assignments, it is beneficial to split large clusters to achieve a more balanced distribution of points across clusters---a desirable characteristic for partition-based indexes used in VSS~\cite{quake, lorann, cuvshierarchical}. 

\vspace*{3mm}
\noindent{\bf Termination Conditions and Final Assignments: } The algorithm terminates after a predetermined number of iterations of STEPS 2 and 3, typically ranging from 5 to 10 for vector embedding datasets~\cite{superkmeans}. Then, STEP 2 is executed once more to obtain the final assignments of each point to its nearest centroid. 
 
\subsection{Efforts to Optimize Vector Clustering}
Research on optimizing clustering of vector embeddings has received far less attention than vector search algorithms. Nevertheless, a few techniques tailored for vector search have seen success when applied to vector clustering~\cite{chen2025vectorchord100m, superkmeans, kmeanselastic, matsui2017pqk, wang2026lindormvector}.

\pagebreak
\noindent{\bf Dimensionality reduction} aims to represent vectors in a lower-dimensional space, thereby reducing memory footprint and compute requirements~\cite{leanvec, li2025saq, happymarriage}.
VectorChord employs a Johnson– Lindenstrauss Transform (JLT)~\cite{jltkmeans} to reduce the dimensionality of the raw vectors before clustering~\cite{chen2025vectorchord100m}. This reduces the memory footprint of clustering while providing performance gains from fewer computations~\cite{zhakubayev2022clustering, jltkmeans, chen2025vectorchord100m}.
LindormVector~\cite{wang2026lindormvector} accomplishes the same goal with a PCA (Principal Component Analysis) projection that concentrates the vectors' ~\textit{energy} in the leading dimensions. More recently, embedding models have been trained to generate Matryoshka embeddings~\cite{kusupati2022matryoshka}, which already concentrate energy in the front dimensions. However, the impact of dimensionality reduction on clustering quality has not been thoroughly studied. 

\vspace*{3mm}\noindent{\bf Quantization} techniques aim to represent each dimension of a vector, or a group of dimensions, with smaller codes while minimizing distortion in the distance metric between points~\cite{rabitq, lvq, surveysystems}. PQk-means is one of the few studies that use quantization prior to clustering~\cite{matsui2017pqk}. PQk-means encodes vectors using Product Quantization~\cite{pqivf}, effectively reducing memory usage and accelerating distance calculations, albeit at the cost of clustering quality. More recent quantization techniques, such as LVQ~\cite{lvq, turbolvq} and RabitQ~\cite{rabitq, rabitqext, gou2025symphonyqg}, have not been used for vector clustering. 

\vspace*{3mm}\noindent{\bf Dimension pruning} aims to break off distance computations when the distance metric has enough resolution to determine that a point will not be the nearest neighbor of a query~\cite{adsampling, pdx, dade, xu2025harmony, dcosurvey1, dcosurvey2}. SuperKMeans integrates dimension pruning into vector clustering by interleaving GEMM routines and pruning kernels, achieving substantial speedups without sacrificing clustering quality~\cite{superkmeans}. However, it remains unclear whether SuperKMeans can be combined with the previously mentioned techniques. 

\vspace*{3mm}
\noindent{\bf The impact of approximations on clustering quality} has not been thoroughly studied. In most clustering pipelines, vectors are indexed at full precision (\codeword{float32}, \codeword{float16}) before being projected into a smaller $d$-dimensional space~\cite{leanvec, gleanvec, li2025saq, happymarriage}, quantized~\cite{rabitqgpu, lvq, leanvec}, and materialized in the index data structure. In this study, we show that approximation techniques can be applied first in the indexing pipeline, saving several round trips to the raw data, reducing memory footprint during clustering, and delivering attractive performance gains, all while mostly preserving clustering quality.

\subsection{Keys for the Performance of Clustering}\label{sec:prelim:resilienc}
GEMM routines for \codeword{float32} are the fuel for high-performance clustering, having undergone decades of optimization. 
Consequently, clustering speed improves with dimensionality reduction, as $d$ decreases while maintaining \codeword{float32} as the data type. Dimension pruning further improves speed by reducing the number of distance calculations. However, the performance of clustering becomes more nuanced when working with quantized vectors. Ideally, the quantization method should facilitate efficient many-to-many distance calculations that leverage modern CPU capabilities for processing smaller data types, such as 8-bit integers. Nonetheless, the performance benefits depend on the availability of SIMD~\cite{kuffo2025bang, numkong, armsme, armvdotq, avx512vpdpbusd} or specialized hardware (e.g., Intel's AMX~\cite{amxintelblog, amxintelpaper}) that can efficiently handle 8-bit data types. 

\begin{figure}[t!]
\centering
\includegraphics[width=1.0\columnwidth]{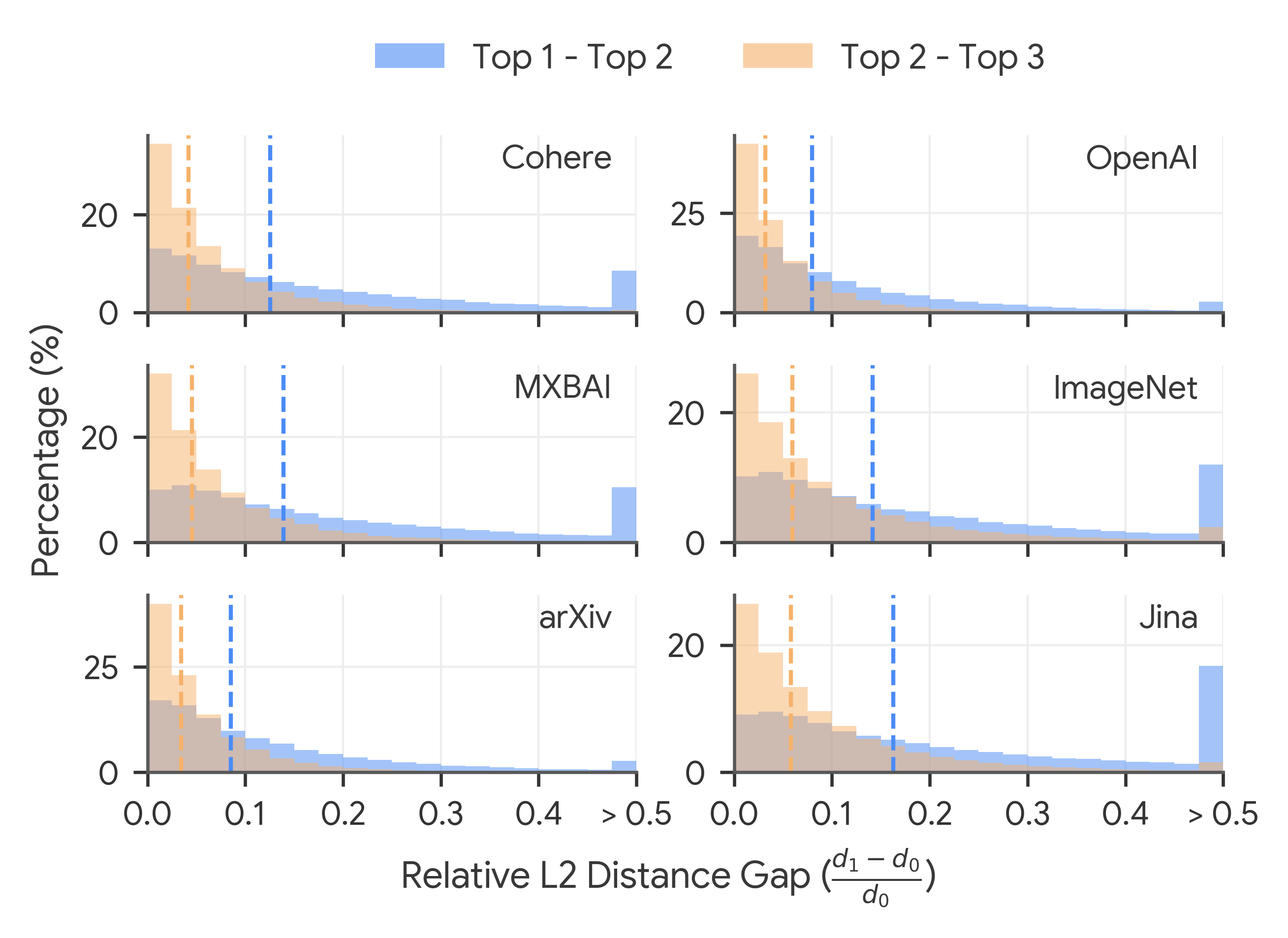}
\vspace*{-8.0mm}
\caption{Distribution of the L2 distance gap between points and their 1st and 2nd nearest centroid (blue) and 2nd and 3rd (yellow) across datasets. The nearest centroid is meaningfully separated, making clustering resilient to approximation techniques. The median is shown as a vertical line.}
\vspace*{-4.0mm}
\label{fig:top}
\end{figure}

\subsection{The Resilience of Clustering}\label{sec:prelim:resilienc}
The assignment step in the clustering process (STEP 2) is effectively a series of \textit{top-1} NNS queries with roles reversed: the vector collection acts as the queries, while the centroids become the vector collection to query. Figure~\ref{fig:top} shows the distribution of the relative L2 distance gap between vectors and their nearest centroids across several vector embedding datasets.
Notably, the gap between the \textit{top-1} and \textit{top-2} nearest centroids (in \colorbox{softblue}{blue}) is 2-3x larger than that of the next neighboring centroids (\textit{top-2} and \textit{top-3}, in \colorbox{lightpeach}{yellow}), making the \textit{top-1} centroid distinctly identifiable. In contrast, the 2nd and 3rd nearest centroids are typically close to one another. This phenomenon makes vector clustering far more resilient to approximation techniques compared to vector search, as errors introduced by methods such as quantization are less likely to affect the correct assignment to the closest centroid~\cite{semanticrecall}. 
This observation raises the question: \textit{To what extent can techniques utilized in vector search be applied to vector clustering?} In the remainder of this study, we empirically address this question by examining the effects of several techniques (and their combinations) on the quality of clustering.

\begin{figure*}[t!]
\centering
\includegraphics[width=0.98\linewidth]{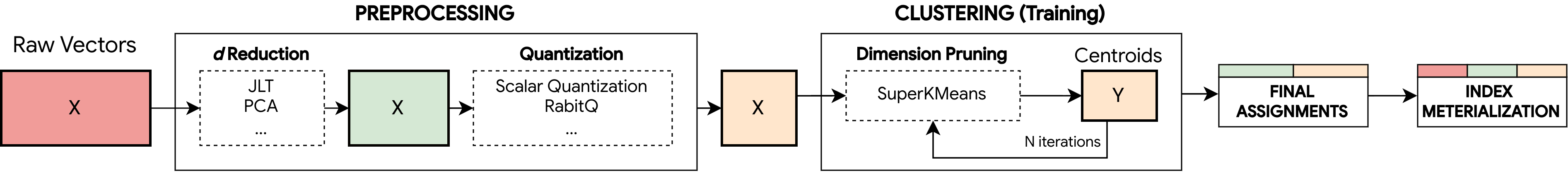}
\vspace*{-4.0mm}
\caption{Our proposed pipeline applies vector approximation techniques before clustering, and dimension pruning accelerates distance calculations during clustering. The final assignment of points to clusters can also happen in the projected (in green) or quantized (in yellow) domain. In our pipeline, touching the raw vectors (in red) after preprocessing is never a must. }
\vspace*{-2.0mm}
\label{fig:pipeline}
\end{figure*}

\begin{table*}[t!]
\renewcommand{\tabcolsep}{2.5pt}
\centering
\caption{Techniques used in our clustering pipeline.}
\vspace*{-4mm}
\resizebox{0.86\linewidth}{!}{%
\begin{tabular}{lclcccc}
\hline
\textbf{Type} & \textbf{Name} & \multicolumn{1}{c}{\textbf{Representation}} & \textbf{\begin{tabular}[c]{@{}c@{}}Centroids \\ Assignment\end{tabular}} & \textbf{\begin{tabular}[c]{@{}c@{}}Updating \\ Centroids\end{tabular}} & \textbf{\begin{tabular}[c]{@{}c@{}}Storage \\ Reduction\end{tabular}} & \multicolumn{1}{c}{\textbf{\begin{tabular}[c]{@{}c@{}}Can be \\ mixed with\end{tabular}}} \\ \hline
\multirow{3}{*}{\textbf{\begin{tabular}[c]{@{}l@{}}Dimensionality\\ Reduction\end{tabular}}} & JLT~\cite{jltkmeans} & \multirow{3}{*}{$x \in \mathbb{R}^{d'}; d' < d$} & \multirow{3}{*}{GEMM} & \multirow{3}{*}{Averaging} & \multirow{3}{*}{$\frac{d}{d'}$x} & \multirow{3}{*}{All except PQ} \\
 & PCA~\cite{pca} &  &  &  &  &  \\ 
 & Matryoshka~\cite{kusupati2022matryoshka} &  &  &  &  &  \\ \hline
\multirow{6}{*}{\textbf{Quantization}} & SQ8 | SQ4~\cite{faisspaper} & $x \mapsto x' \in \{0, 1, ..., 255 | 15\}^{d}$ & 8-bit | 4-bit GEMM & Averaging & 4x | 8x & All \\
 & LVQ4~\cite{lvq} & $x \mapsto x' \in \{0, 1, ..., 15\}^{d}$ & 4-bit GEMM + Adjustment & Fused Decode-Averaging & 8x & All \\
 & RabitQ~\cite{rabitq} & $x \mapsto x' \in \{0, 1\}^{d}$ & FastScan~\cite{originalfastscan} + Adjustment & Fused Decode-Averaging & 30x & All \\
 & PQ8~\cite{matsui2017pqk, pqivf} & $x \mapsto x' \in \{1,...,256\}^{M}$ & FastScan~\cite{originalfastscan} & Sparse voting~\cite{matsui2017pqk} & $\frac{d}{M} \cdot 4$x & None \\
 & PQ4~\cite{matsui2017pqk, pqivf} & $x \mapsto x' \in \{1,...,128\}^{M}$ & FastScan~\cite{originalfastscan} & Sparse voting~\cite{matsui2017pqk} & $\frac{d}{M} \cdot 8$x & None \\ \hline
\textbf{Dimension Pruning} & SuperKMeans~\cite{superkmeans} & Invariant & Invariant + Pruning & Invariant & Invariant & All except PQ \\ \hline
\end{tabular}
}
\label{tab:techniques}
\vspace*{-2mm}
\end{table*}

\section{Our Pipeline: Approximate First, then Cluster}\label{sec:pipeline}

We propose an indexing pipeline (shown in Figure~\ref{fig:pipeline}) where dimensionality reduction and quantization occur \textit{before} clustering, and dimension pruning is used during clustering to reduce the number of distance calculations. This design \textit{differs} from most systems, which typically use full-precision vectors for clustering and apply quantization only after the clustering stage~\cite{lance, milvus, faisspaper, rabitqgpu}. Table~\ref{tab:techniques} presents the techniques we have chosen for our evaluation. Next, we describe how we adapted these techniques for clustering. 


\subsection{Adapting Techniques for Vector Clustering}

\noindent{\textbf{\underline{SQ8 and SQ4:}}} In SQ, each value $x_i$ of a vector $x$ is encoded with a global bias $b = v_{min}$, and a global scaling factor $s = \frac{v_{max} - b}{c_{max}}$, where $c_{max} = 2^B -1 $ and $B$ is the quantization bit-width. Each quantized code $\bar{x}_i$ is defined as $\bar{x}_i = \operatorname{clip} \left( \left\lfloor \frac{x_i - b}{s}\right\rceil, 0, c_{max} \right) $, and reconstructed as: $x_i \approx (\bar{x}_i \cdot s) + b$. Since all vectors are transformed with the same parameters, distance computations can be performed directly in the quantized domain using integer arithmetic, leveraging 8- and 4-bit GEMM routines~\cite{numkong, ruy}. 
Finally, updating centroids can be done by averaging each dimension, also in the quantized domain.

\vspace*{3mm}
\noindent{\textbf{\underline{LVQ4:}}}
In LVQ~\cite{lvq}, each value $x_i$ of a vector $x$ is encoded with a per-vector bias $b_x = x_{min}$, and scale $s_x = \frac{x_{max} - b_x}{c_{max}}; c_{max} = 2^B-1$. Each quantized code $\bar{x}_i$ is defined as $\bar{x}_i = \operatorname{clip} \left( \left\lfloor \frac{x_i - b_x}{s_x}\right\rceil, 0, c_{max} \right) $, and reconstructed as: $x_i \approx (s_x \cdot \bar{x}_i) + b_x$. Since each vector has a different $s$ and $b$, the codes cannot be used directly for distance calculations. To solve this, LVQ proposes fusing reconstruction of \codeword{float32} values with distance calculations. 
However, by doing so we lose the benefits of using a smaller data type. Furthermore, in many-to-many distance calculations, the reconstruction step is performed multiple times for the same vectors, which hurts efficiency. Hereby, we rewrite the L2 distance $||x - y||^2$ as: 

\vspace{-1mm}
\vspace*{-\baselineskip}
\begin{equation}
  \begin{aligned}
||x - y||^2 & \approx \sum_{i}^{d} ((s_x \bar{x}_{i} + b_x) - (s_y \bar{y}_{i} + b_y))^2 \\
          & \approx s_x^2 \sum_{i}^{d} \bar{x}_{i}^2 + s_y^2 \sum_{i}^{d} \bar{y}_{i}^2 - 2 s_x s_y \mathcolorbox{lightpeach}{\langle{\bar{x}, \bar{y}} \rangle} \\ 
          & \;\;\;\; + 2(b_x - b_y) (s_x \sum_{i}^{d} \bar{x}_{i} 
  - s_y \sum_{i}^{d} \bar{y}_{i}) + d (b_x - b_y)^2
\end{aligned}
\end{equation}

\vspace{-1mm}
Collecting constant $x$-only terms into $N_x = s_x^2 \Sigma \bar{x}_{i}^2 + 2 b_x s_x \Sigma \bar{x}_{i} + d b_x^2$, and $y$-only terms into $N_y = s_y^2 \Sigma
  \bar{y}_{i}^2 + 2 b_y s_y \Sigma \bar{y}_{i} + d b_y^2$, and defining $A_x = s_x \Sigma \bar{x} + d b_x$: 

\vspace*{-\baselineskip}
\begin{equation}\label{eq:final:lvq}
  \begin{aligned}
||x - y||^2 \approx N_x + N_y - 2\bigl(s_x \cdot s_y \cdot \mathcolorbox{lightpeach}{\langle \bar{x}, \bar{y} \rangle} + b_y \cdot A_x + b_x \cdot s_y \Sigma \bar{y}_{i}\bigr)
\end{aligned}
\end{equation}

\pagebreak 

This derivation allows us to use 4-bit GEMM kernels to compute the highlighted term. The final distances require additional scalar operations, from which $N_x$ and $A_x$ terms can be cached across $k$-means iterations, while $N_y$ can be computed once at the start of each iteration. Finally, when updating centroids, we cannot solely use LVQ codes, as averaging them is undefined. Thus, we fuse the decoding and averaging of the LVQ codes of the centroid assignments and re-encode the resulting centroids. 

\vspace*{3mm}
\noindent{\textbf{\underline{RabitQ:}}} In RabitQ, each vector is centered, normalized to the unit sphere, and randomly rotated. Then, each dimension is quantized to 1-bit by only preserving their \textit{sign}. Due to the properties of a random rotation, this quantized vector $\bar{x}$ of signs lives in the codebook $\{+1/\sqrt{d},\, -1/\sqrt{d}\}^d$. RabitQ then provides an unbiased estimator of L2 distances and guarantees that the estimator has an asymptotically optimal error bound~\cite{rabitq}. Let $x_r$ be a raw data vector and $y_r$ be a raw query, normalized based on a vector $m$ (i.e., the dataset means). Now, let $x := \frac{x_r - m}{||x_r - m||}$ and $y := \frac{y_r - m}{||y_r - m||}$. The L2 distance between $x_r$ and $y_r$ can be expressed as:

\begin{equation}\label{eq:l2_decomp}
  \begin{aligned}
||x_r - y_r||^2 & = ||(x_r - m) - (y_r - m)||^2 \\
          & = ||x_r - m||^2 + ||y_r - m ||^2 \\
          & \;\;\;\; - 2 \cdot ||x_r - m|| \cdot ||y_r - m|| \cdot \mathcolorbox{softblue}{\langle x, y \rangle} \\ 
\end{aligned}
\end{equation}

\noindent From which $||x_r -m||$ can be computed and cached in the first iteration of $k$-means and $||y_r - m||$ can be precomputed once per centroid, and thus, is amortized by all the pairwise distance calculations. For the term, \mathcolorbox{softblue}{\langle x, y \rangle}, RabitQ derives an unbiased estimator. Let $\bar{x}$ denote the 1-bit quantized version of $x$, then: $\langle x, y \rangle \approx \frac{\langle \bar{x}, y \rangle} {\langle \bar{x}, x \rangle} $, where $\langle \bar{x}, x \rangle$ is the cosine between the original and quantized unit vector, precomputed and stored per data point. Substituting into Equation~\ref{eq:l2_decomp}, and recalling that $y:= \frac{y_r - m}{||y_r - m||}$, then we derive:

\vspace*{-\baselineskip}
\begin{equation}
||x_r - y_r||^2 \approx ||x_r - m||^2 + ||y_r - m||^2
- 2 \cdot \mathcolorbox{softpink}{\frac{||x_r - m||}{\langle \bar{x}, x \rangle}}
\cdot \mathcolorbox{softblue}{\langle \bar{x},\, y_r - m \rangle}
\label{eq:rabitq_l2}
\end{equation}

The scalar factors $||x_r - m||^2$ and $\mathcolorbox{softpink}{||x_r - m||/\langle \bar{x}, x \rangle}$ depend only on the data points and can be cached across $k$-means iterations, while $||y_r - m||^2$ can be computed once at the start of each iteration. In RabitQ, the centered residual vector $y_r - m$ is quantized with SQ4 ($\bar{y}$), with scale $s$ and bias $b$. Then, expanding the SQ4 reconstruction $ (y_r - m)_i \approx \bar{y}_i \cdot s + b$ and reconstructing the codebook 
with the stored bits $\bar{x}_i \in \{0,1\}$, so that each codebook entry is $(2\bar{x}_i- 1)/\sqrt{d}$, yields the following estimator for $\mathcolorbox{softblue}{\langle \bar{x},\, y_r - m \rangle}$:

\vspace*{-\baselineskip}
\begin{equation}\label{eq:final:rabitq}
  \begin{aligned}
\langle \bar{x}, \bar{y} \rangle & \approx \frac{1}{\sqrt{d}} \sum_{i}^{d} (2\bar{x}_{i} - 1)(b + \bar{y}_i \cdot s) \\
          & \approx \frac{1}{\sqrt{d}} \Big[ s \cdot \Big(2 \mathcolorbox{lightpeach}{\sum_{i}^{d} \bar{x}_{i} \cdot \bar{y}_{i}} - \sum_{i}^{d} \bar{y}_i \Big) + b \cdot \Big(\Big(2\sum_{i}^{d}\bar{x_i}\Big) - d \Big) \Big] \\ 
\end{aligned}
\end{equation}

The challenge in computing the highlighted term arises from operand asymmetry: $\bar{x}$ is binary while $\bar{y}$ is SQ4. This computation can be done efficiently using in-register 4-bit lookup tables with the \codeword{PSHUFB}~\cite{pshufb} instruction, which allows for 32 lookups at a time. This method is known as FastScan~\cite{originalfastscan}. The lookup tables can be built on the fly during each iteration of $k$-means. The other terms are scalar operations performed once per distance calculation. In our vector clustering pipeline, we encode data points (which act as queries) with 1-bit RabitQ and the centroids with SQ4. We took this approach because the data points bound the memory footprint of clustering. Finally, when updating centroids, we fuse the decoding of RabitQ codes with the averaging of centroid assignments, then re-encode the resulting centroid with SQ4. 
The decoding of RabitQ codes reconstructs each coordinate $x_{r_i}$ as $x_{r_i} \approx m_i + \frac{1}{\sqrt{d}} \cdot \mathcolorbox{softpink}{||x_r - m||/\langle \bar{x}, x \rangle} \cdot (2\bar{x}_i - 1); \;\; \bar{x}_i \in \{0,1\}$.


\vspace*{3mm}
\noindent{\textbf{\underline{PQ8 and PQ4:}}} In PQ~\cite{pqivf}, the dimensions are divided into $M$, $\frac{d}{M}$ -dimensional subspaces. Each group $M$ is represented with an 8- or 4-bit code from a codebook trained for each subspace. PQ8 allows for 256 different codes, while PQ4 allows for 16 codes. These codes map to representative centroids in each subspace. 
For encoding a vector $x$, each dimension group is assigned the closest centroid code from the codebook of each subspace. As both data and centroids are product-quantized, we use Symmetric Distance Comparisons (code-to-code) to compute assignments. PQ4 can do this efficiently with FastScan~\cite{originalfastscan}, while PQ8 uses scalar lookups as its 8-bit codes are incompatible with the \codeword{PSHUFB} instruction. Finally, when updating centroids, we adopt the \textit{sparse voting} mechanism introduced in PQk-means~\cite{matsui2017pqk}. 


\vspace*{3mm}
\noindent \textbf{\underline{SuperKMeans}} accelerates distance calculations during centroid assignment by interleaving GEMM routines for the front $d'$ dimensions (set to 12.5\% of $d$) and using progressive-pruning kernels every 64 dimensions on the surviving candidates~\cite{superkmeans}. A random orthogonal rotation is required for pruning. This transformation distributes the variance more evenly across all dimensions while preserving the L2 distances between vectors, enabling reliable pruning using ADSampling~\cite{adsampling}. In SQ8 and SQ4, SuperKMeans is extended by replacing the \codeword{float32} GEMM with 8- and 4-bit GEMMs. However, SuperKMeans encounters limitations with LVQ and RabitQ because it requires L2 distances at $d'$, where $d' < d$. Both LVQ and RabitQ derivations (Equations~\ref{eq:final:lvq} and ~\ref{eq:l2_decomp}) can be used with $d'$, as the L2 distance can be decomposed as $||x-y||^2 = ||x'-y'||^2 + ||x''-y''||^2$, where $x'$ refers to the front $d'$ dimensions of $x$, and $x''$ to the remaining ones. However, this incurs storage overhead for the adjustment factors needed per dimension segment, as well as the additional overhead of floating-point operations. Hereby, for LVQ4 and RabitQ, we use only 2 pruning checkpoints at 12.5\% and 25\% of $d$. For progressive pruning to take place, efficient 1-to-1 distance kernels are also necessary. In SQ8, SQ4, and LVQ4, we use 8- and 4-bit L2 distance kernels~\cite{numkong}. In RabitQ, we use the \codeword{POPCNT}-per-bitplane approach described in ~\cite{rabitq}. 

\vspace*{3mm}
\noindent \textbf{\underline{The Johnson--Lindenstrauss Transform (JLT)}} projects vectors to a lower $d'$ while preserving pairwise distances between the points~\cite{jltlemma, jltkmeans}. \textbf{\underline{Principal Component Analysis (PCA)}} projects vectors into a lower $d'$ while preserving global variance and concentrating the \textit{energy} of the vectors in the front dimensions.  \textbf{\underline{Matryoshka}} vectors are produced by embedding models that already concentrate the energy in the front dimensions~\cite{kusupati2022matryoshka}. These techniques do not need extra engineering to be integrated in the clustering pipeline, as vectors remain in the \codeword{float32} domain.

\begin{table}[t]
\renewcommand{\tabcolsep}{2.5pt}
\centering
\caption{Vector embedding datasets used for evaluation}
\vspace*{-4mm}
\label{tab:datasets}
\resizebox{1.0\columnwidth}{!}{%
\begin{tabular}{lllccccc}
\hline
\multicolumn{1}{l}{\textbf{Name}} & \multicolumn{1}{l}{\textbf{Type}} & \textbf{Model} & \textbf{\# Vectors} & \textbf{Dim.} & \multicolumn{1}{c}{\textbf{Size (GB)$\uparrow$}} & \multicolumn{1}{c}{\textbf{LID}} & \multicolumn{1}{c}{\textbf{Dist.}} \\ \hline
Cohere~\cite{datasetcohere}    & Text  & EmbedV3-EN     & 10,000,000 & 1024 & 40.96 & 27.9 & \raisebox{-.4\height}{\includegraphics[width=0.04\textwidth]{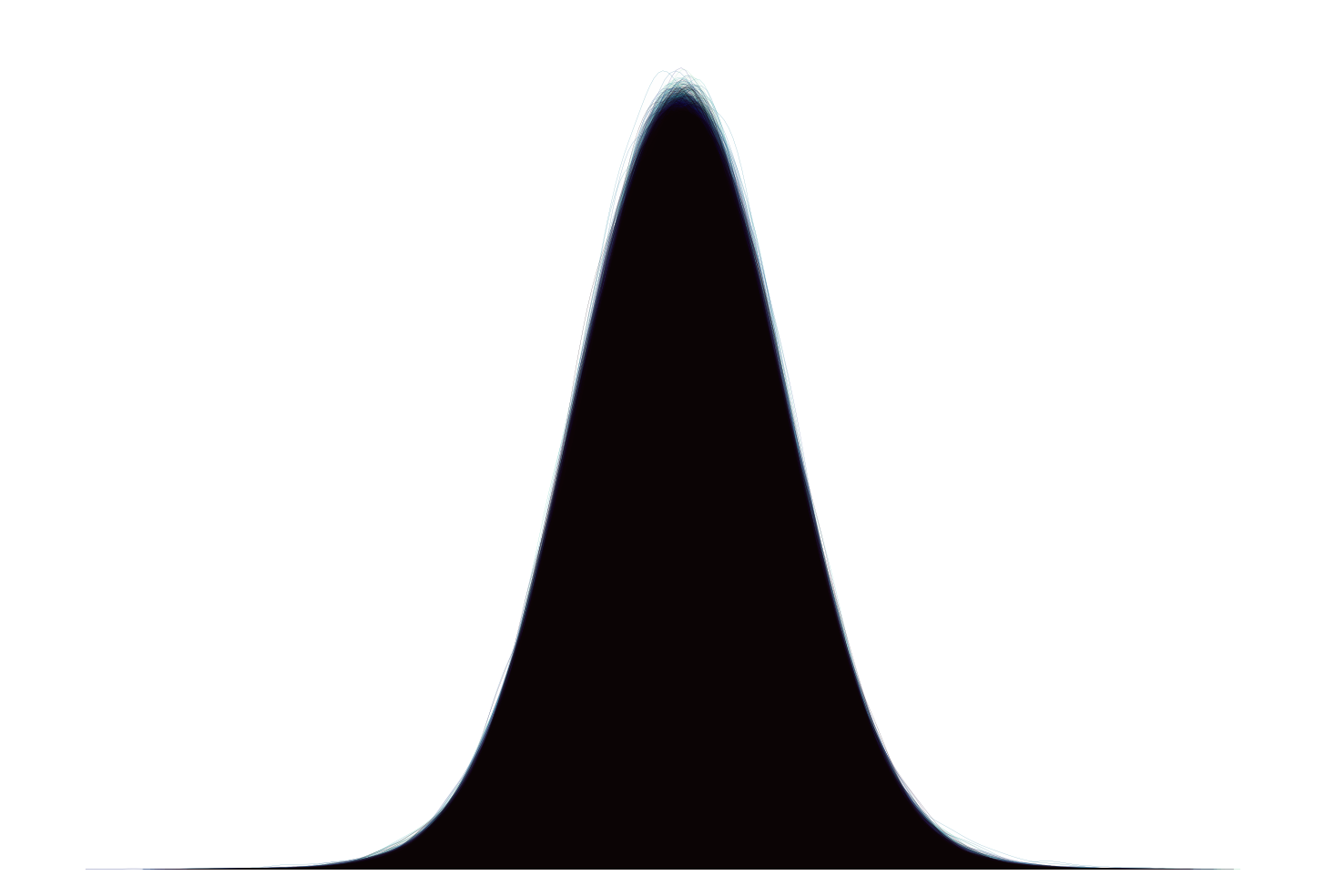}} \\
OpenAI~\cite{datasetopenai}    & Text  & OpenAI         & 5,000,000    & 1536 & 30.72 & 31.4 & \raisebox{-.4\height}{\includegraphics[width=0.04\textwidth]{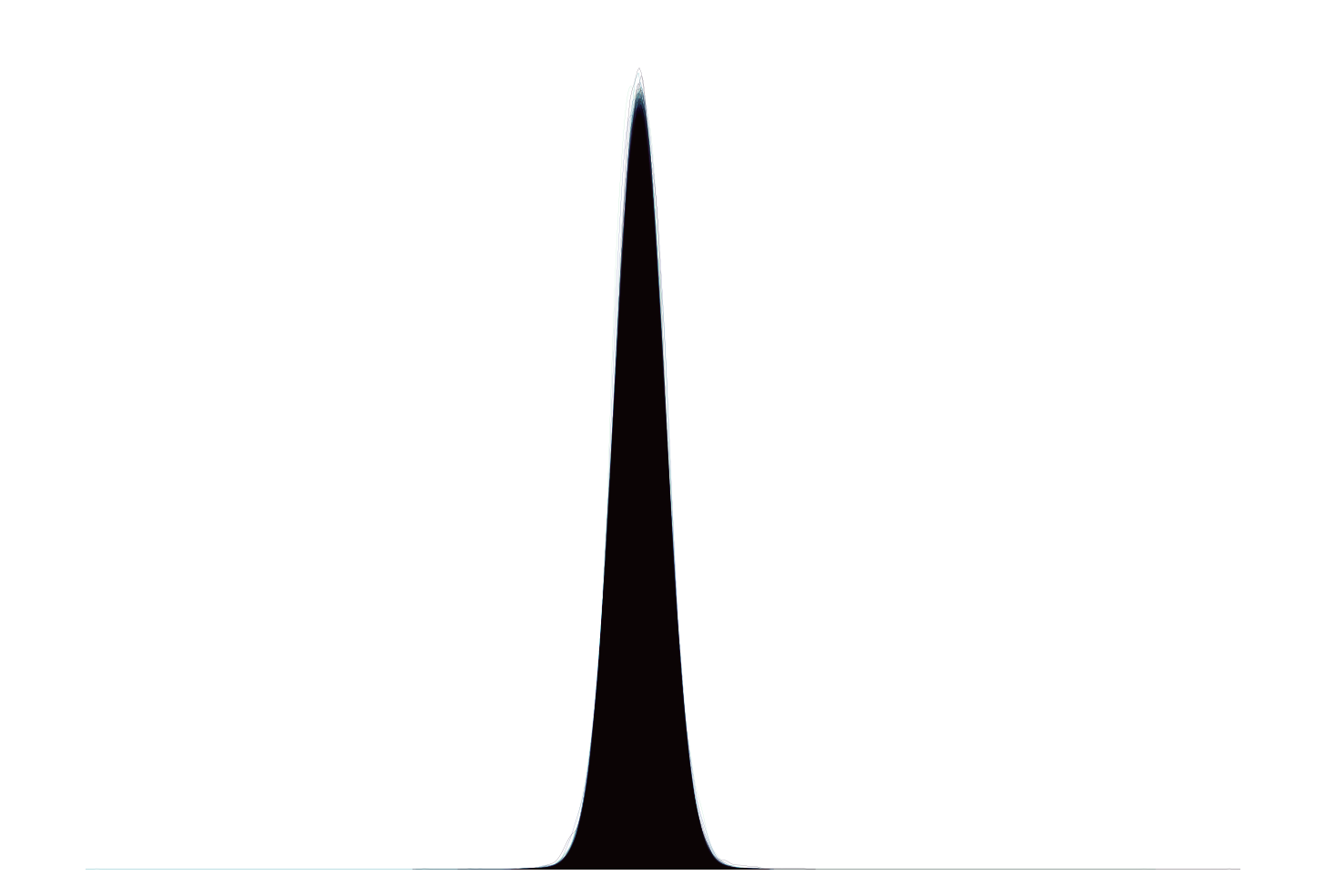}} \\
arXiv~\cite{datasetarxiv}     & Text  & InstructorXL   & 2,253,000  & 768  & 6.92 & 28.9 & \raisebox{-.4\height}{\includegraphics[width=0.04\textwidth]{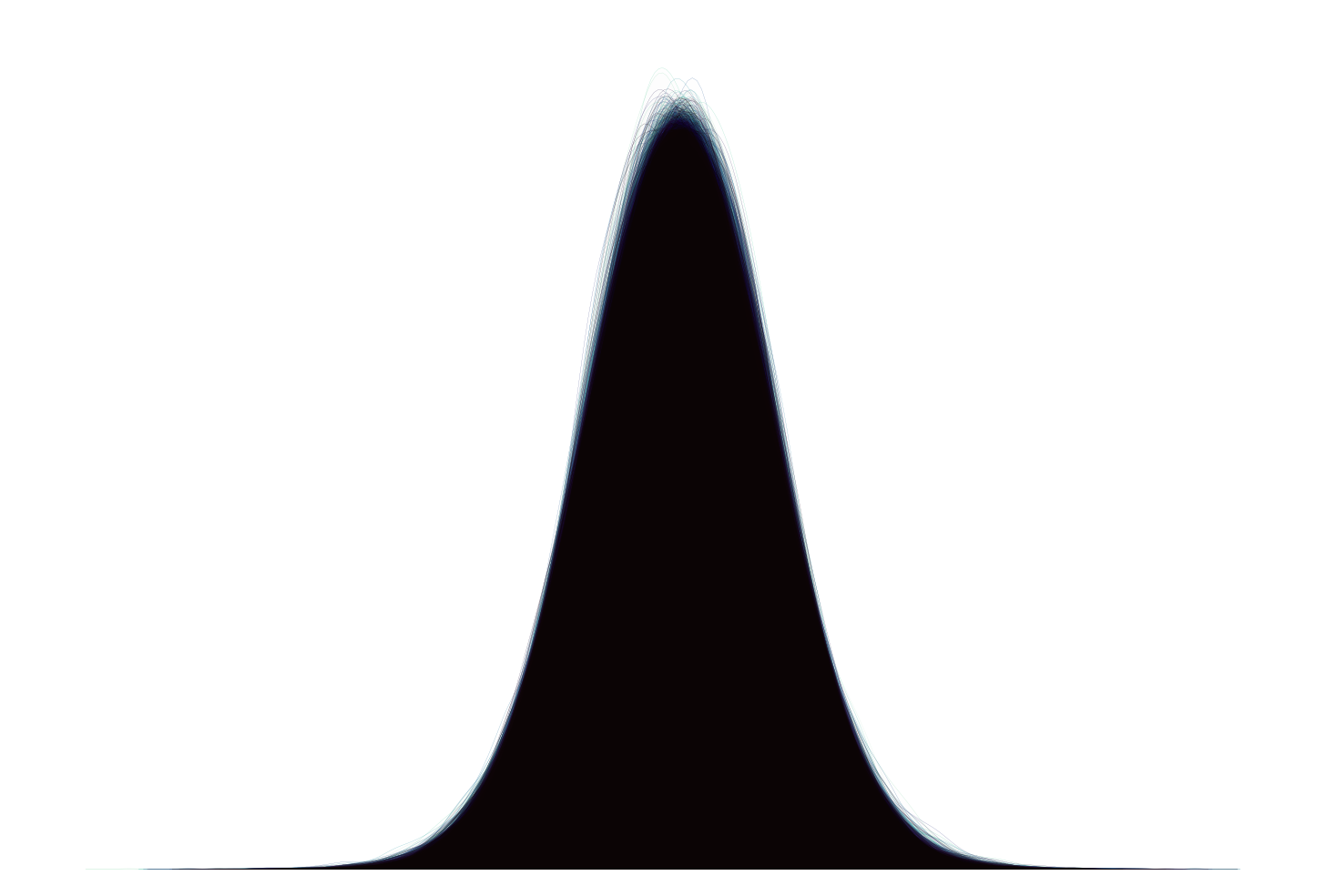}} \\
Jina~\cite{datasetjina, vibe}  & Code & Jina           & 1,374,067  & 768  & 4.22 & 20.5 & \raisebox{-.4\height}{\includegraphics[width=0.04\textwidth]{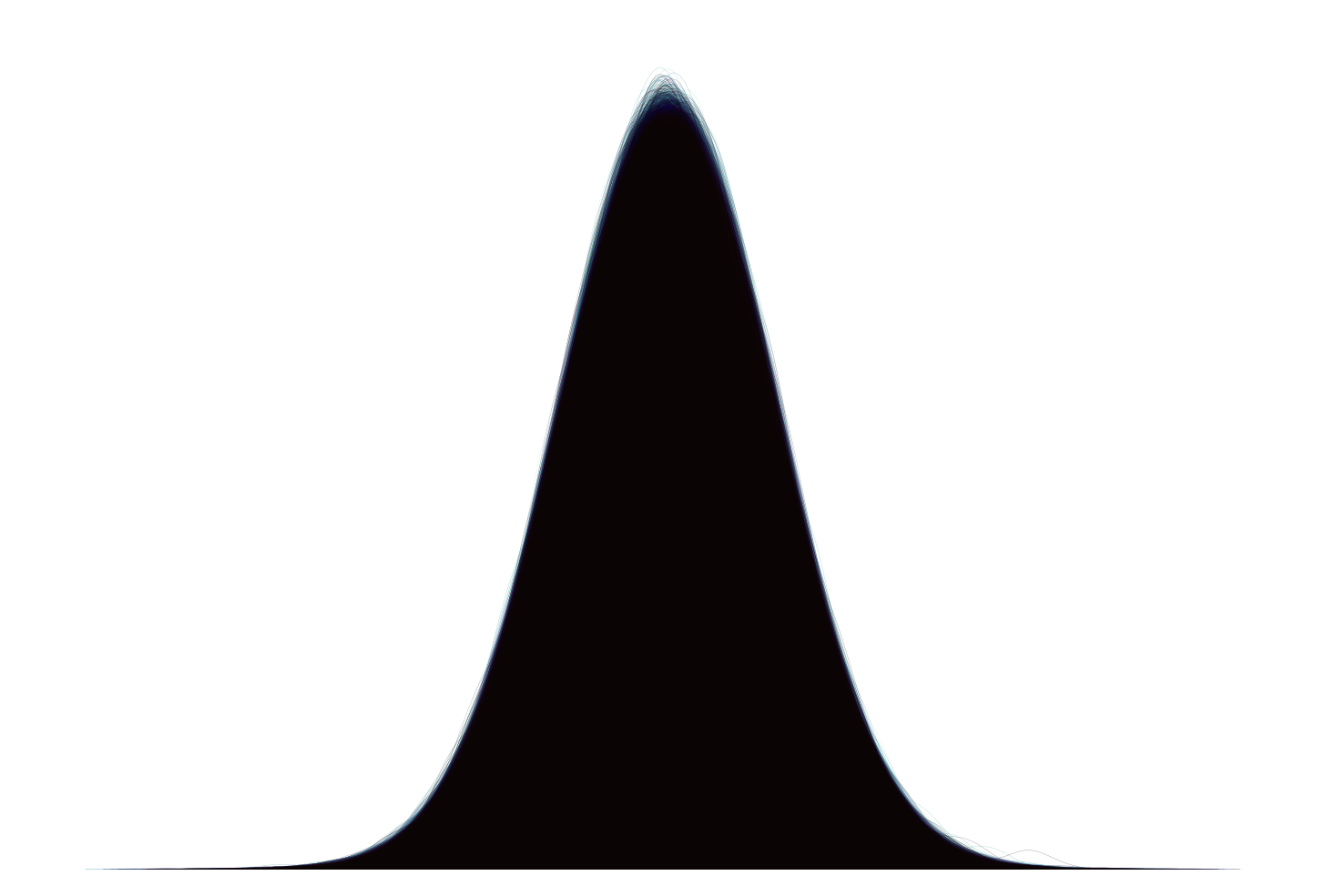}} \\
MXBAI~\cite{vibe, mxbai}     & Text  & MXBAI          & 769,382    & 1024 & 3.15 & 24.5 & \raisebox{-.4\height}{\includegraphics[width=0.04\textwidth]{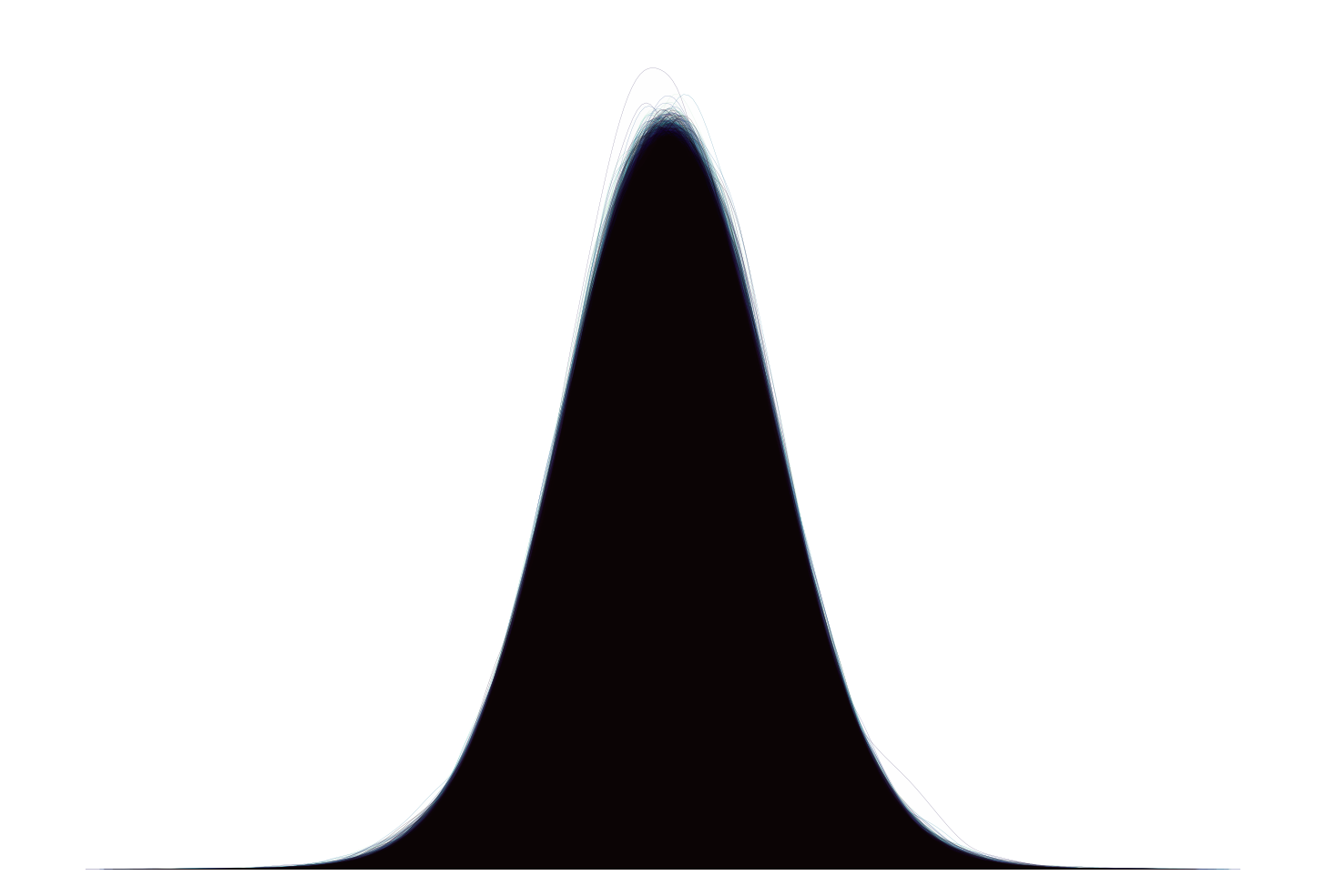}} \\
ImageNet~\cite{vibe, deng2009imagenet} & Image & CLIP           & 1,281,167  & 512  & 2.62 & 14.6 &  \raisebox{-.4\height}{\includegraphics[width=0.04\textwidth]{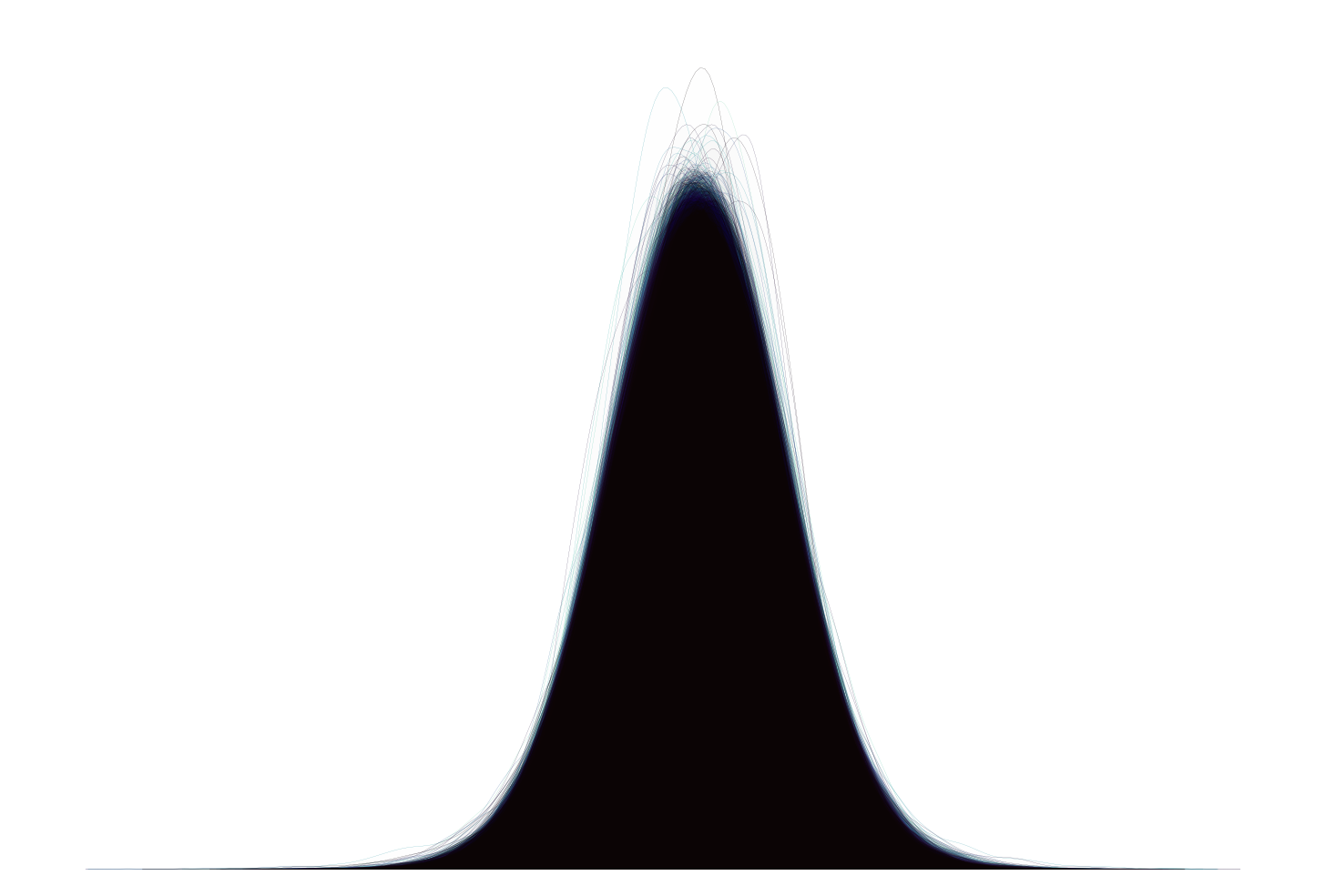}} \\
\hline

\end{tabular}
}
\vspace*{-2mm}
\end{table}


\section{Evaluation}\label{sec:eval}

We experimentally evaluate our indexing pipeline using the vector embedding datasets presented in Table~\ref{tab:datasets}. Our evaluation focuses on three aspects: (i) clustering quality, (ii) storage reduction, and (iii) clustering speed. For clustering quality, we assess how well the generated centroids perform in vector search tasks when used as an entry point for an IVF index. To quantify this, we use the \textit{recall@k} metric together with the number of vectors explored during search, which serves as a proxy for the amount of distance computation required.  \textit{Recall@k} measures the proportion of the true nearest neighbors (i.e., the ground truth) that are retrieved among the top-$k$ results. In our evaluation, we retrieve the top 100 neighbors (i.e., \textit{recall@100}). In IVF indexes, \textit{recall} can be tuned by varying the number of clusters probed during search. We report results for two probing configurations that correspond to exploring 1\% and 3\% of the available clusters. For a given \textit{recall} level, exploring fewer vectors is preferable as it indicates that the search requires fewer distance computations.

Additionally, we record the \textit{within-cluster sum of squares} (WCSS), which measures the compactness of clusters by summing the squared distances of each point to its centroid (lower is better). Regarding storage reduction, we report the reduction in storage required to perform the clustering iterations. Finally, in terms of clustering speed, we measure the end-to-end runtime of our clustering pipeline over 10 iterations of $k$-means, setting the number of clusters to $4\sqrt{N}$~\cite{faisspaper, milvus}, where $N$ is the number of vectors in the collection.

\begin{figure*}[t!]
\centering
\includegraphics[width=1.0\linewidth]{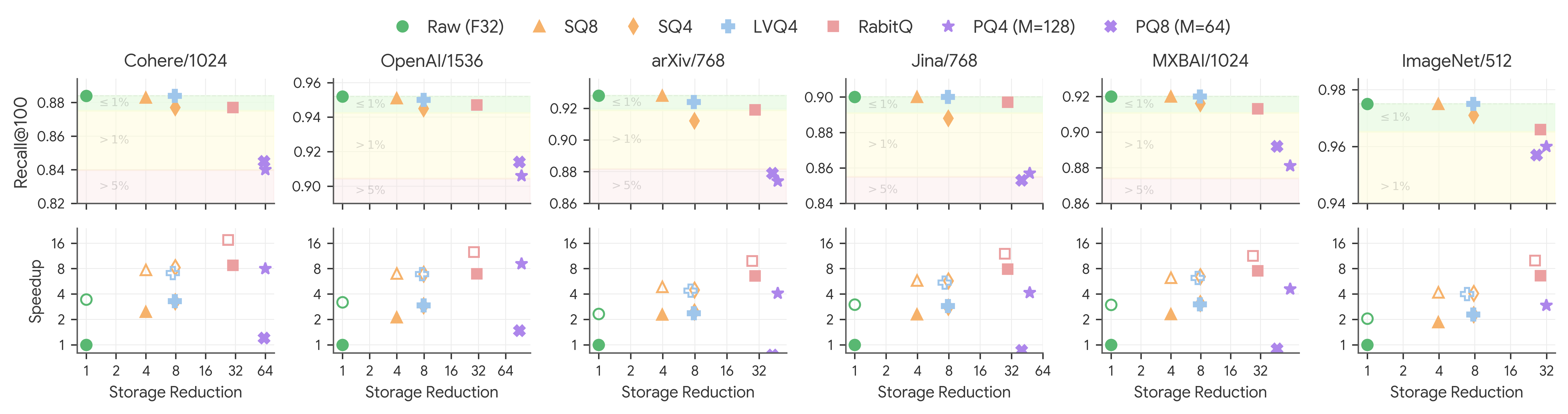}
\vspace*{-8.0mm}
\caption{The effect on clustering quality (top) and speedup (bottom) of quantizing vectors before clustering. SQ8, LVQ4, and RabitQ achieve near-optimal clustering quality (within 1\% of ideal). SuperKMeans (shown with hollow markers) accelerates clustering up to 17x. SuperKMeans is not shown in the top plot since recall is unaffected. 
\colorbox[HTML]{eafbe5}{Green}, \colorbox[HTML]{fffddb}{yellow}, and \colorbox[HTML]{f9e0e0}{red} indicate $\leq$1\%, $>$1\%, and $>$5\% away, resp., from the results of clustering raw vectors. 
}
\vspace*{-3.0mm}
\label{fig:eval:quantization}
\end{figure*}

\begin{table}[H]
\renewcommand{\tabcolsep}{1.5pt}
\centering
\caption{Quality of the generated centroids for VSS tasks when clustering quantized vectors. To measure recall, we do top-100 IVF index searches by probing 1\% and 3\% of the clusters. \textit{Color coding is the same as in Figure~\ref{fig:eval:quantization}.}}
\vspace*{-4mm}
\label{tab:eval:quantization}
\resizebox{1.0\columnwidth}{!}{%
\begin{tabular}{lcccccccc}
\hline
\textbf{Quant.} & \textbf{\begin{tabular}[c]{@{}c@{}}Build \\ Time (s)\end{tabular}} & \textbf{\begin{tabular}[c]{@{}c@{}}Build \\ Time (s) \\ w/ SKM\end{tabular}} & \textbf{\begin{tabular}[c]{@{}c@{}}Recall\\ @100\\ @1\%\end{tabular}} & \textbf{\begin{tabular}[c]{@{}c@{}}Vectors \\ explored\\ @1\% ($\times 10^{3}$)\end{tabular}} & \textbf{\begin{tabular}[c]{@{}c@{}}Recall\\ @100\\ @3\%\end{tabular}} & \textbf{\begin{tabular}[c]{@{}c@{}}Vectors \\ explored\\ @3\% ($\times 10^{3}$)\end{tabular}} & \textbf{\begin{tabular}[c]{@{}c@{}}WCSS \\ ($\times 10^{6}$)\end{tabular}} & \textbf{\begin{tabular}[c]{@{}c@{}}Cluster \\ Size\\ Std. Dev.\end{tabular}} \\ \hline
\rowcolor[HTML]{EFEFEF} 
\multicolumn{9}{c}{\cellcolor[HTML]{EFEFEF}\textbf{Cohere ($N=10M$, $d=1024$, $k=12649$)}} \\
\rowcolor[HTML]{EFEFEF} 
\textbf{Raw} & 640.3 (1.0x) & 186.2 (3.4x) & 0.884 & 105.43 & 0.934 & 313.32 & 5.103 & 354 \\
\textbf{SQ8} & 259.3 (2.5x) & 82.5 (7.8x) & \cellcolor[HTML]{EAFBE5}0.883 & \cellcolor[HTML]{EAFBE5}105.45 & \cellcolor[HTML]{EAFBE5}0.933 & \cellcolor[HTML]{EAFBE5}313.23 & \cellcolor[HTML]{EAFBE5}5.106 & \cellcolor[HTML]{EAFBE5}352 \\
\textbf{SQ4} & 198.1 (3.2x) & 77.4 (8.3x) & \cellcolor[HTML]{EAFBE5}0.877 & \cellcolor[HTML]{EAFBE5}103.68 & \cellcolor[HTML]{EAFBE5}0.930 & \cellcolor[HTML]{EAFBE5}308.60 & \cellcolor[HTML]{F9E0E0}5.697 & \cellcolor[HTML]{EAFBE5}345 \\
\textbf{LVQ4} & 194.8 (3.3x) & 90.0 (7.1x) & \cellcolor[HTML]{EAFBE5}0.884 & \cellcolor[HTML]{FFFDDB}106.69 & \cellcolor[HTML]{EAFBE5}0.934 & \cellcolor[HTML]{FFFDDB}317.50 & \cellcolor[HTML]{EAFBE5}5.106 & \cellcolor[HTML]{EAFBE5}353 \\
\textbf{RabitQ} & \textbf{73.3 (8.7x)} & \textbf{36.5 (17.5x)} & \cellcolor[HTML]{EAFBE5}0.877 & \cellcolor[HTML]{FFFDDB}107.31 & \cellcolor[HTML]{EAFBE5}0.930 & \cellcolor[HTML]{FFFDDB}319.11 & \cellcolor[HTML]{FFFDDB}5.289 & \cellcolor[HTML]{EAFBE5}348 \\
\textbf{PQ4} & 80.5 (8.0x) & - & \cellcolor[HTML]{FFFDDB}0.840 & \cellcolor[HTML]{EAFBE5}98.01 & \cellcolor[HTML]{FFFDDB}0.906 & \cellcolor[HTML]{EAFBE5}290.62 & \cellcolor[HTML]{F9E0E0}6.694 & \cellcolor[HTML]{EAFBE5}315 \\
\textbf{PQ8} & 532.8 (1.2x) & - & \cellcolor[HTML]{FFFDDB}0.845 & \cellcolor[HTML]{EAFBE5}97.38 & \cellcolor[HTML]{FFFDDB}0.908 & \cellcolor[HTML]{EAFBE5}288.74 & \cellcolor[HTML]{F9E0E0}6.640 & \cellcolor[HTML]{EAFBE5}323 \\ \hline
\rowcolor[HTML]{EFEFEF} 
\multicolumn{9}{c}{\cellcolor[HTML]{EFEFEF}\textbf{OpenAI ($N=5M$, $d=1536$, $k=8944$)}} \\
\rowcolor[HTML]{EFEFEF} 
\textbf{Raw} & 284.8 (1.0x) & 89.0 (3.2x) & 0.952 & 51.86 & 0.982 & 153.73 & 0.993 & 216 \\
\textbf{SQ8} & 134.6 (2.1x) & 40.7 (7.0x) & \cellcolor[HTML]{EAFBE5}0.951 & \cellcolor[HTML]{EAFBE5}51.87 & \cellcolor[HTML]{EAFBE5}0.982 & \cellcolor[HTML]{EAFBE5}153.75 & \cellcolor[HTML]{EAFBE5}0.994 & \cellcolor[HTML]{EAFBE5}216 \\
\textbf{SQ4} & 98.4 (2.9x) & 41.2 (6.9x) & \cellcolor[HTML]{EAFBE5}0.945 & \cellcolor[HTML]{EAFBE5}51.45 & \cellcolor[HTML]{EAFBE5}0.978 & \cellcolor[HTML]{EAFBE5}152.48 & \cellcolor[HTML]{F9E0E0}1.200 & \cellcolor[HTML]{EAFBE5}211 \\
\textbf{LVQ4} & 97.3 (2.9x) & 41.3 (6.9x) & \cellcolor[HTML]{EAFBE5}0.950 & \cellcolor[HTML]{EAFBE5}52.29 & \cellcolor[HTML]{EAFBE5}0.982 & \cellcolor[HTML]{EAFBE5}155.42 & \cellcolor[HTML]{EAFBE5}0.995 & \cellcolor[HTML]{EAFBE5}217 \\
\textbf{RabitQ} & \textbf{41.0 (6.9x)} & \textbf{22.6 (12.6x)} & \cellcolor[HTML]{EAFBE5}0.947 & \cellcolor[HTML]{EAFBE5}52.20 & \cellcolor[HTML]{EAFBE5}0.979 & \cellcolor[HTML]{EAFBE5}155.03 & \cellcolor[HTML]{FFFDDB}1.014 & \cellcolor[HTML]{EAFBE5}218 \\
\textbf{PQ4} & 31.3 (9.1x) & - & \cellcolor[HTML]{FFFDDB}0.906 & \cellcolor[HTML]{EAFBE5}51.93 & \cellcolor[HTML]{EAFBE5}0.964 & \cellcolor[HTML]{EAFBE5}152.51 & \cellcolor[HTML]{F9E0E0}1.432 & \cellcolor[HTML]{EAFBE5}208 \\
\textbf{PQ8} & 193.5 (1.5x) & - & \cellcolor[HTML]{FFFDDB}0.914 & \cellcolor[HTML]{FFFDDB}54.28 & \cellcolor[HTML]{EAFBE5}0.967 & \cellcolor[HTML]{EAFBE5}152.11 & \cellcolor[HTML]{F9E0E0}1.361 & \cellcolor[HTML]{F9E0E0}299 \\ \hline
\rowcolor[HTML]{EFEFEF} 
\multicolumn{9}{c}{\cellcolor[HTML]{EFEFEF}\textbf{arXiv ($N=2.25M$, $d=768$, $k=6003$)}} \\
\rowcolor[HTML]{EFEFEF} 
\textbf{Raw} & 42.9 (1.0x) & 18.5 (2.3x) & 0.928 & 24.09 & 0.980 & 71.17 & 0.384 & 140 \\
\textbf{SQ8} & 18.7 (2.3x) & 8.8 (4.9x) & \cellcolor[HTML]{EAFBE5}0.928 & \cellcolor[HTML]{EAFBE5}24.04 & \cellcolor[HTML]{EAFBE5}0.979 & \cellcolor[HTML]{EAFBE5}71.09 & \cellcolor[HTML]{EAFBE5}0.385 & \cellcolor[HTML]{EAFBE5}140 \\
\textbf{SQ4} & 17.6 (2.4x) & 9.6 (4.5x) & \cellcolor[HTML]{FFFDDB}0.912 & \cellcolor[HTML]{EAFBE5}24.29 & \cellcolor[HTML]{EAFBE5}0.973 & \cellcolor[HTML]{EAFBE5}71.69 & \cellcolor[HTML]{F9E0E0}0.485 & \cellcolor[HTML]{EAFBE5}140 \\
\textbf{LVQ4} & 18.1 (2.4x) & 9.8 (4.4x) & \cellcolor[HTML]{EAFBE5}0.924 & \cellcolor[HTML]{EAFBE5}24.06 & \cellcolor[HTML]{EAFBE5}0.978 & \cellcolor[HTML]{EAFBE5}71.07 & \cellcolor[HTML]{EAFBE5}0.386 & \cellcolor[HTML]{FFFDDB}142 \\
\textbf{RabitQ} & \textbf{6.6 (6.5x)} & \textbf{4.3 (9.9x)} & \cellcolor[HTML]{EAFBE5}0.919 & \cellcolor[HTML]{EAFBE5}24.27 & \cellcolor[HTML]{EAFBE5}0.976 & \cellcolor[HTML]{EAFBE5}71.77 & \cellcolor[HTML]{FFFDDB}0.401 & \cellcolor[HTML]{FFFDDB}143 \\
\textbf{PQ4} & 10.5 (4.1x) & - & \cellcolor[HTML]{F9E0E0}0.874 & \cellcolor[HTML]{FFFDDB}24.48 & \cellcolor[HTML]{EAFBE5}0.959 & \cellcolor[HTML]{EAFBE5}71.04 & \cellcolor[HTML]{F9E0E0}0.556 & \cellcolor[HTML]{EAFBE5}141 \\
\textbf{PQ8} & 56.6 (0.8x) & - & \cellcolor[HTML]{F9E0E0}0.879 & \cellcolor[HTML]{EAFBE5}24.35 & \cellcolor[HTML]{EAFBE5}0.960 & \cellcolor[HTML]{EAFBE5}69.27 & \cellcolor[HTML]{F9E0E0}0.538 & \cellcolor[HTML]{F9E0E0}161 \\ \hline
\rowcolor[HTML]{EFEFEF} 
\multicolumn{9}{c}{\cellcolor[HTML]{EFEFEF}\textbf{Jina ($N=1.37M$, $d=768$, $k=4688$)}} \\
\rowcolor[HTML]{EFEFEF} 
\textbf{Raw} & 26.0 (1.0x) & 8.7 (3.0x) & 0.900 & 14.19 & 0.952 & 42.17 & 0.992 & 135 \\
\textbf{SQ8} & 11.4 (2.3x) & 4.5 (5.8x) & \cellcolor[HTML]{EAFBE5}0.901 & \cellcolor[HTML]{EAFBE5}14.21 & \cellcolor[HTML]{EAFBE5}0.953 & \cellcolor[HTML]{EAFBE5}42.18 & \cellcolor[HTML]{EAFBE5}0.992 & \cellcolor[HTML]{EAFBE5}135 \\
\textbf{SQ4} & 9.3 (2.8x) & 4.5 (5.7x) & \cellcolor[HTML]{FFFDDB}0.888 & \cellcolor[HTML]{F9E0E0}15.46 & \cellcolor[HTML]{EAFBE5}0.943 & \cellcolor[HTML]{FFFDDB}44.27 & \cellcolor[HTML]{F9E0E0}1.138 & \cellcolor[HTML]{F9E0E0}295 \\
\textbf{LVQ4} & 9.1 (2.9x) & 4.8 (5.5x) & \cellcolor[HTML]{EAFBE5}0.901 & \cellcolor[HTML]{EAFBE5}14.22 & \cellcolor[HTML]{EAFBE5}0.952 & \cellcolor[HTML]{EAFBE5}42.25 & \cellcolor[HTML]{EAFBE5}0.992 & \cellcolor[HTML]{EAFBE5}136 \\
\textbf{RabitQ} & \textbf{3.3 (7.9x)} & \textbf{2.2 (12.0x)} & \cellcolor[HTML]{EAFBE5}0.897 & \cellcolor[HTML]{EAFBE5}14.23 & \cellcolor[HTML]{EAFBE5}0.950 & \cellcolor[HTML]{EAFBE5}42.20 & \cellcolor[HTML]{FFFDDB}1.039 & \cellcolor[HTML]{FFFDDB}137 \\
\textbf{PQ4} & 6.3 (4.2x) & - & \cellcolor[HTML]{FFFDDB}0.857 & \cellcolor[HTML]{FFFDDB}14.37 & \cellcolor[HTML]{EAFBE5}0.928 & \cellcolor[HTML]{FFFDDB}42.64 & \cellcolor[HTML]{F9E0E0}1.496 & \cellcolor[HTML]{EAFBE5}124 \\
\textbf{PQ8} & 30.2 (0.9x) & - & \cellcolor[HTML]{F9E0E0}0.853 & \cellcolor[HTML]{FFFDDB}14.75 & \cellcolor[HTML]{EAFBE5}0.924 & \cellcolor[HTML]{FFFDDB}43.11 & \cellcolor[HTML]{F9E0E0}1.488 & \cellcolor[HTML]{F9E0E0}176 \\ \hline
\rowcolor[HTML]{EFEFEF} 
\multicolumn{9}{c}{\cellcolor[HTML]{EFEFEF}\textbf{MXBAI ($N=769K$, $d=1024$, $k=3508$)}} \\
\rowcolor[HTML]{EFEFEF} 
\textbf{Raw} & 14.3 (1.0x) & 4.8 (3.0x) & 0.920 & 8.60 & 0.970 & 25.25 & 101.334 & 97 \\
\textbf{SQ8} & 6.2 (2.3x) & 2.3 (6.3x) & \cellcolor[HTML]{EAFBE5}0.920 & \cellcolor[HTML]{EAFBE5}8.62 & \cellcolor[HTML]{EAFBE5}0.970 & \cellcolor[HTML]{EAFBE5}25.31 & \cellcolor[HTML]{EAFBE5}101.361 & \cellcolor[HTML]{EAFBE5}97 \\
\textbf{SQ4} & 4.6 (3.1x) & 2.2 (6.5x) & \cellcolor[HTML]{EAFBE5}0.916 & \cellcolor[HTML]{EAFBE5}8.56 & \cellcolor[HTML]{EAFBE5}0.968 & \cellcolor[HTML]{EAFBE5}25.12 & \cellcolor[HTML]{F9E0E0}113.845 & \cellcolor[HTML]{EAFBE5}95 \\
\textbf{LVQ4} & 4.7 (3.0x) & 2.3 (6.2x) & \cellcolor[HTML]{EAFBE5}0.922 & \cellcolor[HTML]{FFFDDB}8.72 & \cellcolor[HTML]{EAFBE5}0.971 & \cellcolor[HTML]{FFFDDB}25.68 & \cellcolor[HTML]{EAFBE5}101.447 & \cellcolor[HTML]{FFFDDB}99 \\
\textbf{RabitQ} & \textbf{1.9 (7.5x)} & \textbf{1.3 (11.3x)} & \cellcolor[HTML]{EAFBE5}0.913 & \cellcolor[HTML]{FFFDDB}8.86 & \cellcolor[HTML]{EAFBE5}0.967 & \cellcolor[HTML]{FFFDDB}26.04 & \cellcolor[HTML]{FFFDDB}105.737 & \cellcolor[HTML]{FFFDDB}101 \\
\textbf{PQ4} & 3.1 (4.6x) & - & \cellcolor[HTML]{FFFDDB}0.881 & \cellcolor[HTML]{FFFDDB}8.33 & \cellcolor[HTML]{EAFBE5}0.953 & \cellcolor[HTML]{EAFBE5}24.12 & \cellcolor[HTML]{F9E0E0}151.481 & \cellcolor[HTML]{EAFBE5}88 \\
\textbf{PQ8} & 15.9 (0.9x) & - & \cellcolor[HTML]{FFFDDB}0.892 & \cellcolor[HTML]{FFFDDB}8.35 & \cellcolor[HTML]{EAFBE5}0.956 & \cellcolor[HTML]{EAFBE5}23.53 & \cellcolor[HTML]{F9E0E0}141.776 & \cellcolor[HTML]{F9E0E0}105 \\ \hline
\rowcolor[HTML]{EFEFEF} 
\multicolumn{9}{c}{\cellcolor[HTML]{EFEFEF}\textit{\textbf{ImageNet ($N=1.28M$, $d=512$, $k=4527$)}}} \\
\rowcolor[HTML]{EFEFEF} 
\textbf{Raw} & 14.3 (1.0x) & 7.0 (2.0x) & 0.975 & 12.86 & 0.995 & 37.86 & 0.293 & 110 \\
\textbf{SQ8} & 7.7 (1.9x) & 3.4 (4.2x) & \cellcolor[HTML]{EAFBE5}0.976 & \cellcolor[HTML]{EAFBE5}12.88 & \cellcolor[HTML]{EAFBE5}0.995 & \cellcolor[HTML]{EAFBE5}37.89 & \cellcolor[HTML]{EAFBE5}0.293 & \cellcolor[HTML]{EAFBE5}110 \\
\textbf{SQ4} & 6.3 (2.3x) & 3.5 (4.1x) & \cellcolor[HTML]{EAFBE5}0.971 & \cellcolor[HTML]{EAFBE5}12.92 & \cellcolor[HTML]{EAFBE5}0.994 & \cellcolor[HTML]{EAFBE5}37.99 & \cellcolor[HTML]{F9E0E0}0.345 & \cellcolor[HTML]{FFFDDB}112 \\
\textbf{LVQ4} & 6.2 (2.3x) & 3.6 (4.0x) & \cellcolor[HTML]{EAFBE5}0.975 & \cellcolor[HTML]{EAFBE5}12.93 & \cellcolor[HTML]{EAFBE5}0.996 & \cellcolor[HTML]{EAFBE5}38.13 & \cellcolor[HTML]{EAFBE5}0.294 & \cellcolor[HTML]{EAFBE5}109 \\
\textbf{RabitQ} & \textbf{2.2 (6.6x)} & \textbf{1.4 (10.0x)} & \cellcolor[HTML]{EAFBE5}0.966 & \cellcolor[HTML]{EAFBE5}12.81 & \cellcolor[HTML]{EAFBE5}0.993 & \cellcolor[HTML]{EAFBE5}37.82 & \cellcolor[HTML]{F9E0E0}0.326 & \cellcolor[HTML]{EAFBE5}111 \\
\textbf{PQ4} & 4.9 (2.9x) & - & \cellcolor[HTML]{FFFDDB}0.960 & \cellcolor[HTML]{FFFDDB}13.05 & \cellcolor[HTML]{EAFBE5}0.992 & \cellcolor[HTML]{EAFBE5}37.89 & \cellcolor[HTML]{F9E0E0}0.418 & \cellcolor[HTML]{EAFBE5}106 \\
\textbf{PQ8} & 23.4 (0.6x) & - & \cellcolor[HTML]{FFFDDB}0.957 & \cellcolor[HTML]{EAFBE5}12.90 & \cellcolor[HTML]{EAFBE5}0.990 & \cellcolor[HTML]{EAFBE5}37.19 & \cellcolor[HTML]{F9E0E0}0.399 & \cellcolor[HTML]{FFFDDB}113 \\ \hline
\end{tabular}
}
\vspace*{-2mm}
\end{table}

\pagebreak

\noindent{\bf Index Materialization: } The last step of our pipeline materializes the index to be used for VSS. Notably, the same representations can be used for both VSS and clustering, which avoids a round-trip to the raw vectors during index materialization. Recent studies have proposed vector indexes that utilize not only quantization but also dimensionality reduction~\cite{happymarriage, leanvec, gleanvec, li2025saq}, aligning with the principles of our pipeline. To ensure a fair comparison across techniques, we compute the \textit{recall@k} based on cluster membership. In other words, we assume that the vectors within the probed clusters are ranked correctly. This helps us circumvent artifacts introduced by VSS pipelines, such as re-ranking. Finally, we materialize the trained centroids for the IVF index as \codeword{float32} vectors. Unless stated otherwise, these centroids are computed solely from the reconstruction of the quantized codes of the trained centroids. 

\vspace*{3mm}
\noindent{\bf Hardware and Software:} We used an AMD Zen 5 EPYC 9R45 CPU (4.5 GHz) with 256GB of RAM and 32 cores (\codeword{r8a.8xlarge} in AWS). Our implementations extend the C++ SuperKMeans codebase~\cite{superkmeanscode}. We use 8- and 4-bit GEMM kernels found in the NumKong library~\cite{numkong} and \codeword{float32} GEMM kernels from OpenBLAS~\cite{openblas}. Our experiments use all the available cores. 

\begin{figure*}[t!]
\centering
\includegraphics[width=1.0\linewidth]{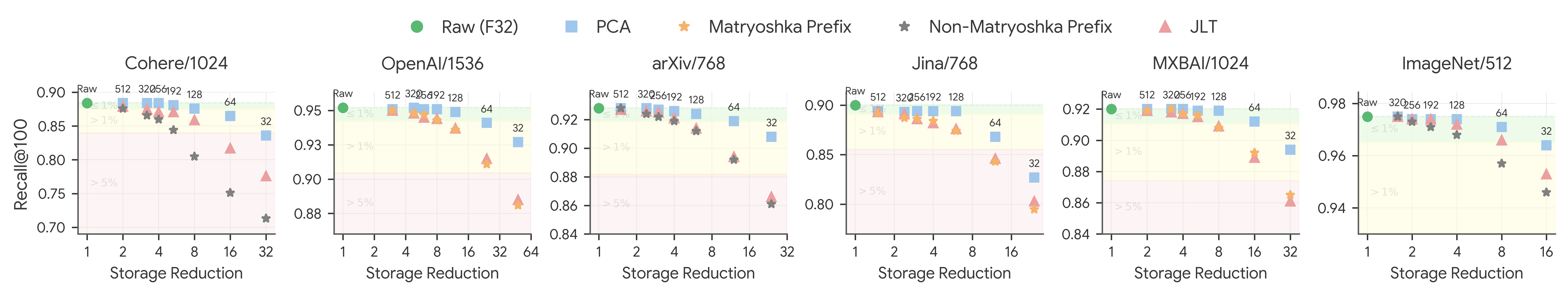}
\vspace*{-8.0mm}
\caption{The effect of reducing vector dimensionality on clustering quality. PCA is the best option to maintain the quality of centroids. Dimensionality is shown as labels. \textit{Color coding is the same as in Figure~\ref{fig:eval:quantization}.} 
}
\vspace*{-3.0mm}
\label{fig:eval:dimreduction}
\end{figure*}

\begin{figure*}[t!]
\centering
\includegraphics[width=1.0\linewidth]{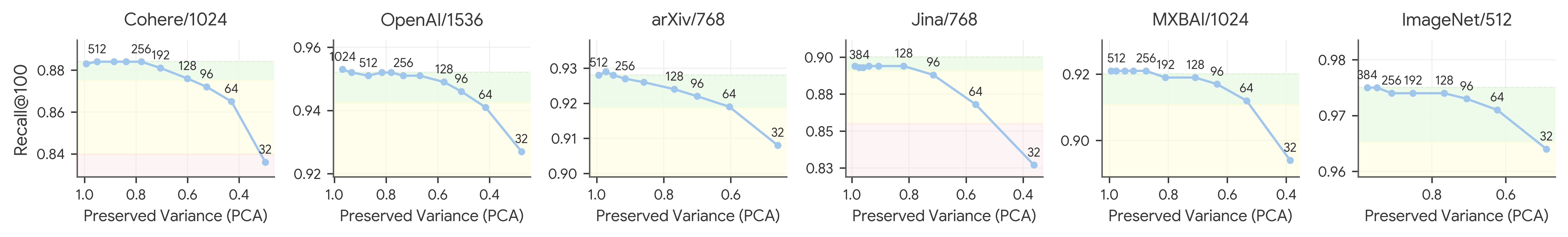}
\vspace*{-8.0mm}
\caption{Around 70\% of preserved variance after the PCA projection is sufficient to achieve near-optimal clustering quality (within 1\%). \textit{Color coding is the same as in Figure~\ref{fig:eval:quantization}}. }
\vspace*{-3.0mm}
\label{fig:eval:variance}
\end{figure*}

\subsection{Quantization Techniques}\label{sec:eval:quantization}

Figure~\ref{fig:eval:quantization} (top) shows the quality of the resulting centroids for vector search tasks based on the recall they yield when used as entry points for an IVF index and probing 1\% of clusters. Table~\ref{tab:eval:quantization} presents these results in detail, reporting several metrics that characterize clustering quality. \textbf{SQ8} always achieves a clustering quality on-par with using raw ~\codeword{float32} vectors in all aspects, while providing 4x storage reduction and up to 8x speedups when paired with SuperKMeans (bottom of Figure~\ref{fig:eval:quantization}). \textbf{LVQ4} and \textbf{RabitQ} achieve near-optimal clustering quality (with no more than 1\% difference in recall) with 8x and 30x storage reduction, resp., while mostly maintaining cluster balance and providing significant speedups. RabitQ results in a slightly higher imbalance in cluster size. However, it provides up to 17x acceleration when paired with SuperKMeans' pruning, where the additional metadata to use SuperKMeans degrades storage reduction by around 10\%. Finally, LVQ4 outperforms \textbf{SQ4} in terms of clustering quality, despite both achieving the same level of storage reduction.


\begin{table}[h]
\renewcommand{\tabcolsep}{3.0pt}
\centering
\caption{Profiling of clustering in the Cohere dataset. The additional work of quantization methods (encoding and precomputing terms) is negligible, except on PQ4 and PQ8. The phases of SuperKMeans (SKM) are broken down. }
\vspace*{-4mm}
\label{tab:profiling}
\resizebox{1.0\columnwidth}{!}{%
\begin{tabular}{lllllllll}
\hline
\textbf{Stage} & \multicolumn{1}{c}{\textbf{Raw}} & \multicolumn{1}{c}{\textbf{+SKM}} & \multicolumn{1}{c}{\textbf{SQ8}} & \multicolumn{1}{c}{\textbf{SQ4}} & \multicolumn{1}{c}{\textbf{LVQ4}} & \multicolumn{1}{c}{\textbf{RabitQ}} & \multicolumn{1}{c}{\textbf{PQ4}} & \multicolumn{1}{c}{\textbf{PQ8}} \\ \hline
Assignments (10 iters.) & 635.6 & 177.3 & 76.4 & 71.6 & 84.3 & 30.9 & 52.9 & 470.9 \\
- 1st Iteration (SuperKMeans) & - & - 31\% & - 33\% & - 26\% & - 23\% & - 22\% & - & - \\
- Front $d'$ (SuperKMeans) & - & - 45\% & - 35\% & - 43\% & - 51\% & - 51\% & - & - \\
- Pruning (SuperKMeans) & - & - 24\% & - 32\% & - 33\% & - 29\% & - 28\% & - & - \\ \hline
Precomputing Terms & - & - & - & - & 0.1 & 0.1 & - & - \\ \hline
Encoding & - & - & 0.6 & 0.5 & 0.3 & 0.4 & 26.3 & 59.3 \\ \hline
Random Rotation & - & 3.2 & 3.3 & 3.3 & 3.2 & 3.3 & - & - \\ \hline
Updating Centroids & 2.3 & 2.3 & 0.8 & 0.6 & 0.7 & 0.6 & 0.8 & 2.1 \\ \hline
Other & 2.3 & 3.3 & 1.4 & 1.3 & 1.5 & 1.1 & 0.5 & 0.5 \\ \hline \hline
Total & 640.3 & 186.2 & 82.5 & 77.4 & 90 & 36.5 & 80.5 & 532.8 \\ \hline
\end{tabular}
}
\vspace*{-4mm}
\end{table}

On the other hand, \textbf{PQ} vectors provide the highest storage reduction but compromise clustering quality, affecting both recall and the balance of clusters. PQ struggles with clustering quality because it builds a global codebook from raw vectors, whereas PQ is typically applied to the residuals after clustering, allowing vectors to be centered around the clusters' means~\cite{faisspaper}. A way to improve the quality of PQ-based clustering is to use the raw vectors to update the centroids during the final $k$-means iteration. This brings the WCSS into the \colorbox[HTML]{fffddb}{yellow} zone, improving recall from 0.84 to 0.85 in the Cohere dataset, albeit at the cost of accessing the raw vectors. 

Table~\ref{tab:profiling} breaks down the clustering time into different phases. All techniques, except for PQ4 and PQ8, use SuperKMeans. Notably, the time spent on encoding and precomputing constant terms in SQ, LVQ, and RabitQ is negligible. In contrast, encoding consumes a significant portion of PQ's runtime, limiting its speedup. Finally, in RabitQ and LVQ, a considerable amount of time during the assignment step is consumed by the overhead of computing partial L2 distances of the front $d'$ dimensions.


\vspace*{3mm}
\noindent{\bf Closing RabitQ's Gap in Balance:} Despite RabitQ achieving exceptional end-to-end recall, the balance of clusters is negatively impacted. Further inspection reveals that when clustering RabitQ codes, outer clusters covering the periphery of the data distribution bleed boundary points into denser in-distribution clusters--explaining the higher number of vectors explored. These outer cluster centroids are the furthest from the dataset mean ($m$ in Equation~\ref{eq:rabitq_l2}), which is also the root of the issue. Recall that the decoding of RabitQ codes reconstructs each coordinate $x_{r_i}$ as $x_{r_i} \approx m_i + \frac{1}{\sqrt{d}} \cdot \mathcolorbox{softpink}{||x_r - m||/\langle \bar{x}, x \rangle} \cdot (2\bar{x}_i - 1); \;\; \bar{x}_i \in \{0,1\}$. Hereby, the decoded residual sits \textit{on} the sphere of radius \mathcolorbox{softpink}{||x_r - m||/\langle \bar{x}, x \rangle} around $m$. By Cauchy–Schwarz, $||x_r - m||_1 \le \sqrt{d} \; ||x_r - m||$, so this radius is always at least $||x_r - m||$. Consequently, each RabitQ-decoded vector has a systematic outward reconstruction bias from the dataset mean. When these decoded vectors are averaged to form a centroid, the bias largely cancels for inner clusters (whose constituent sign patterns are diverse) but survives for peripheral clusters (whose vectors' sign patterns are highly aligned). The result is that peripheral cluster centroids are reconstructed less precisely than central ones, overshooting outward. A simple way to counteract this effect is to use the raw vectors in the last iteration to update the centroids. This brings RabitQ clusters into the \colorbox{lightgreen}{green} zone across all metrics, albeit at the cost of accessing the raw vectors, which may be unfeasible if they reside in a slower storage unit. 

\subsection{Dimensionality Reduction Techniques}\label{sec:eval:dimreduction}
Figure~\ref{fig:eval:dimreduction} shows the quality of the resulting centroids for vector search tasks when dimensionality reduction techniques are applied to the vectors prior to clustering. Table~\ref{tab:eval:dimred} presents these results in detail. \textbf{PCA} is the most effective method for preserving the quality of the centroids. Figure~\ref{fig:eval:variance} shows that preserving 60--70\% of the total variance after applying PCA is sufficient to achieve clustering quality within 1\% of the ideal, while maintaining 80\% of the variance (3-4x reduction of $d$) achieves optimal quality. In contrast, the quality of \textbf{JLT} projections degrades much sooner. Using \textbf{Matryoshka} prefixes performs comparably to JLT on the OpenAI, Jina, and MXBAI datasets, whose vectors stem from a model that produces Matryoshka representations. 

\begin{figure*}[t!]
\centering
\includegraphics[width=1.0\linewidth]{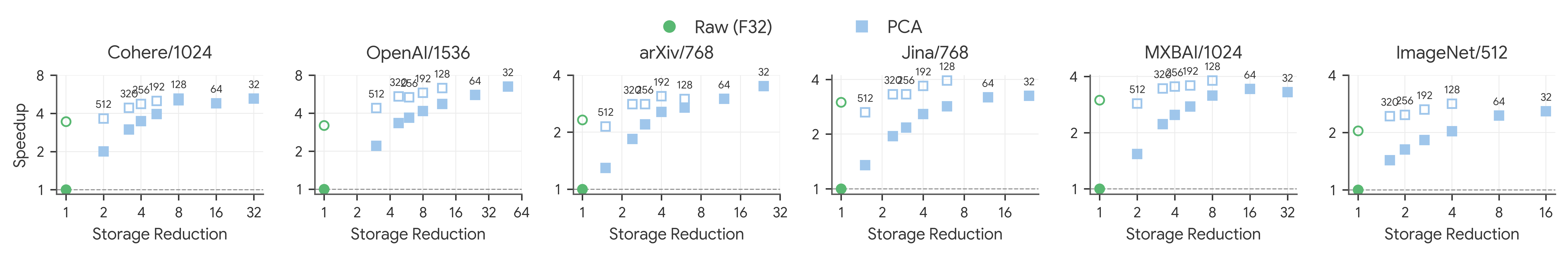}
\vspace*{-8.0mm}
\caption{The effect of PCA projections on clustering speed. Speedup stops improving with thinner vectors. However, SuperKMeans makes wider vectors have comparable speedup to those of thinner vectors without the loss in quality. }
\vspace*{-3.0mm}
\label{fig:eval:dimreduction:speed}
\end{figure*}

\begin{table}[H]
\renewcommand{\tabcolsep}{1.5pt}
\centering
\caption{ Quality of the generated centroids for VSS tasks as vector dimensionality is reduced. PCA projections achieve the highest quality. Only the datasets with Matryoshka embeddings are shown. }
\vspace*{-4mm}
\label{tab:eval:dimred}
\resizebox{1.0\columnwidth}{!}{%
\begin{tabular}{lcccccccc}
\hline
\textbf{Tech.} & \textbf{\begin{tabular}[c]{@{}c@{}}Build \\ Time (s)\end{tabular}} & \textbf{\begin{tabular}[c]{@{}c@{}}Build \\ Time (s) \\ w/ SKM\end{tabular}} & \textbf{\begin{tabular}[c]{@{}c@{}}Prep.\\ Time\\ (s)\end{tabular}} & \textbf{\begin{tabular}[c]{@{}c@{}}Recall\\ @100\\ @1\%\end{tabular}} & \textbf{\begin{tabular}[c]{@{}c@{}}Vectors \\ explored\\ @1\% ($\times 10^{3}$)\end{tabular}} & \textbf{\begin{tabular}[c]{@{}c@{}}Recall\\ @100\\ @3\%\end{tabular}} & \textbf{\begin{tabular}[c]{@{}c@{}}WCSS \\ ($\times 10^{6}$)\end{tabular}} & \textbf{\begin{tabular}[c]{@{}c@{}}Cluster \\ Size\\ Std. Dev.\end{tabular}} \\ \hline
\rowcolor[HTML]{EFEFEF} 
\multicolumn{9}{c}{\cellcolor[HTML]{EFEFEF}\textbf{OpenAI ($N=5M$, $d=1536$, $k=8944$)}} \\
\rowcolor[HTML]{EFEFEF} 
Raw & 284.8 (1.0x) & 89.0 (3.2x) & - & 0.952 & 51.86 & 0.982 & 0.993 & 216 \\
PCA 512 & 129.1 (2.2x) & 64.9 (4.4x) & 8.6 & \cellcolor[HTML]{EAFBE5}0.951 & \cellcolor[HTML]{EAFBE5}51.65 & \cellcolor[HTML]{EAFBE5}0.982 & \cellcolor[HTML]{EAFBE5}0.993 & \cellcolor[HTML]{EAFBE5}210 \\
PCA 256 & 77.3 (3.7x) & 53.2 (5.3x) & 7.1 & \cellcolor[HTML]{EAFBE5}0.951 & \cellcolor[HTML]{EAFBE5}50.86 & \cellcolor[HTML]{EAFBE5}0.982 & \cellcolor[HTML]{EAFBE5}0.994 & \cellcolor[HTML]{EAFBE5}204 \\
PCA 128 & 60.2 (4.7x) & 45.1 (6.3x) & 6.4 & \cellcolor[HTML]{EAFBE5}0.949 & \cellcolor[HTML]{EAFBE5}49.34 & \cellcolor[HTML]{EAFBE5}0.981 & \cellcolor[HTML]{EAFBE5}0.999 & \cellcolor[HTML]{EAFBE5}197 \\
PCA 64 & 51.0 (5.6x) & 45.1 (6.3x) & 6.1 & \cellcolor[HTML]{FFFDDB}0.941 & \cellcolor[HTML]{EAFBE5}46.87 & \cellcolor[HTML]{EAFBE5}0.977 & \cellcolor[HTML]{FFFDDB}1.013 & \cellcolor[HTML]{EAFBE5}203 \\
PCA 32 & 43.8 (6.5x) & 37.0 (7.7x) & 6.2 & \cellcolor[HTML]{FFFDDB}0.927 & \cellcolor[HTML]{EAFBE5}45.57 & \cellcolor[HTML]{EAFBE5}0.973 & \cellcolor[HTML]{FFFDDB}1.039 & \cellcolor[HTML]{F9E0E0}239 \\ \hline
JLT 512 & 96.1 (3.0x) & 49.6 (5.7x) & 3.2 & \cellcolor[HTML]{EAFBE5}0.950 & \cellcolor[HTML]{EAFBE5}51.29 & \cellcolor[HTML]{EAFBE5}0.981 & \cellcolor[HTML]{EAFBE5}0.995 & \cellcolor[HTML]{EAFBE5}208 \\
JLT 256 & 55.5 (5.1x) & 42.4 (6.7x) & 1.8 & \cellcolor[HTML]{EAFBE5}0.945 & \cellcolor[HTML]{EAFBE5}49.67 & \cellcolor[HTML]{EAFBE5}0.979 & \cellcolor[HTML]{EAFBE5}0.997 & \cellcolor[HTML]{EAFBE5}206 \\
JLT 128 & 40.2 (7.1x) & 37.9 (7.5x) & 1.1 & \cellcolor[HTML]{FFFDDB}0.937 & \cellcolor[HTML]{EAFBE5}45.43 & \cellcolor[HTML]{EAFBE5}0.973 & \cellcolor[HTML]{FFFDDB}1.006 & \cellcolor[HTML]{EAFBE5}213 \\
JLT 64 & 30.6 (9.3x) & 32.8 (8.7x) & 0.7 & \cellcolor[HTML]{FFFDDB}0.915 & \cellcolor[HTML]{EAFBE5}38.98 & \cellcolor[HTML]{FFFDDB}0.958 & \cellcolor[HTML]{FFFDDB}1.033 & \cellcolor[HTML]{F9E0E0}320 \\
JLT 32 & 32.8 (8.7x) & 30.5 (9.3x) & 0.7 & \cellcolor[HTML]{F9E0E0}0.885 & \cellcolor[HTML]{EAFBE5}44.14 & \cellcolor[HTML]{FFFDDB}0.935 & \cellcolor[HTML]{F9E0E0}1.113 & \cellcolor[HTML]{F9E0E0}657 \\ \hline
Mat. 512 & 104.3 (2.7x) & 57.8 (4.9x) & - & \cellcolor[HTML]{EAFBE5}0.950 & \cellcolor[HTML]{EAFBE5}51.39 & \cellcolor[HTML]{EAFBE5}0.981 & \cellcolor[HTML]{EAFBE5}0.994 & \cellcolor[HTML]{EAFBE5}210 \\
Mat. 256 & 63.5 (4.5x) & 54.0 (5.3x) & - & \cellcolor[HTML]{EAFBE5}0.947 & \cellcolor[HTML]{EAFBE5}49.53 & \cellcolor[HTML]{EAFBE5}0.980 & \cellcolor[HTML]{EAFBE5}0.998 & \cellcolor[HTML]{EAFBE5}206 \\
Mat. 128 & 45.8 (6.2x) & 50.9 (5.6x) & - & \cellcolor[HTML]{FFFDDB}0.936 & \cellcolor[HTML]{EAFBE5}45.44 & \cellcolor[HTML]{EAFBE5}0.973 & \cellcolor[HTML]{FFFDDB}1.006 & \cellcolor[HTML]{EAFBE5}218 \\
Mat. 64 & 37.6 (7.6x) & 45.1 (6.3x) & - & \cellcolor[HTML]{FFFDDB}0.911 & \cellcolor[HTML]{EAFBE5}38.84 & \cellcolor[HTML]{FFFDDB}0.954 & \cellcolor[HTML]{FFFDDB}1.035 & \cellcolor[HTML]{F9E0E0}330 \\
Mat. 32 & 39.0 (7.3x) & 45.6 (6.2x) & - & \cellcolor[HTML]{F9E0E0}0.881 & \cellcolor[HTML]{EAFBE5}45.47 & \cellcolor[HTML]{F9E0E0}0.930 & \cellcolor[HTML]{F9E0E0}1.12 & \cellcolor[HTML]{F9E0E0}693 \\ \hline
\rowcolor[HTML]{EFEFEF} 
\multicolumn{9}{c}{\cellcolor[HTML]{EFEFEF}\textbf{Jina ($N=1.37M$, $d=768$, $k=4688$)}} \\
\rowcolor[HTML]{EFEFEF} 
Raw & 26.0 (1.0x) & 8.7 (3.0x) & - & 0.900 & 14.19 & 0.952 & 0.992 & 135 \\
PCA 512 & 19.3 (1.4x) & 9.9 (2.6x) & 2.0 & \cellcolor[HTML]{EAFBE5}0.894 & \cellcolor[HTML]{EAFBE5}13.95 & \cellcolor[HTML]{EAFBE5}0.949 & \cellcolor[HTML]{EAFBE5}0.806 & \cellcolor[HTML]{EAFBE5}132 \\
PCA 256 & 12.0 (2.2x) & 7.9 (3.3x) & 1.6 & \cellcolor[HTML]{EAFBE5}0.894 & \cellcolor[HTML]{EAFBE5}13.87 & \cellcolor[HTML]{EAFBE5}0.949 & \cellcolor[HTML]{EAFBE5}0.807 & \cellcolor[HTML]{EAFBE5}130 \\
PCA 128 & 9.2 (2.8x) & 6.6 (3.9x) & 1.4 & \cellcolor[HTML]{EAFBE5}0.894 & \cellcolor[HTML]{EAFBE5}13.76 & \cellcolor[HTML]{EAFBE5}0.949 & \cellcolor[HTML]{EAFBE5}0.808 & \cellcolor[HTML]{EAFBE5}132 \\
PCA 64 & 8.2 (3.2x) & 6.2 (4.2x) & 1.4 & \cellcolor[HTML]{FFFDDB}0.868 & \cellcolor[HTML]{EAFBE5}12.79 & \cellcolor[HTML]{FFFDDB}0.934 & \cellcolor[HTML]{EAFBE5}0.824 & \cellcolor[HTML]{EAFBE5}123 \\
PCA 32 & 8.0 (3.2x) & 5.8 (4.5x) & 1.6 & \cellcolor[HTML]{F9E0E0}0.827 & \cellcolor[HTML]{EAFBE5}12.29 & \cellcolor[HTML]{FFFDDB}0.909 & \cellcolor[HTML]{EAFBE5}0.866 & \cellcolor[HTML]{EAFBE5}133 \\ \hline
JLT 512 & 17.2 (1.5x) & 8.8 (3.0x) & 0.8 & \cellcolor[HTML]{EAFBE5}0.893 & \cellcolor[HTML]{EAFBE5}13.87 & \cellcolor[HTML]{EAFBE5}0.948 & \cellcolor[HTML]{EAFBE5}0.807 & \cellcolor[HTML]{EAFBE5}131 \\
JLT 256 & 10.0 (2.6x) & 7.4 (3.5x) & 0.4 & \cellcolor[HTML]{FFFDDB}0.886 & \cellcolor[HTML]{EAFBE5}13.53 & \cellcolor[HTML]{EAFBE5}0.943 & \cellcolor[HTML]{EAFBE5}0.811 & \cellcolor[HTML]{EAFBE5}131 \\
JLT 128 & 7.4 (3.5x) & 7.1 (3.7x) & 0.3 & \cellcolor[HTML]{FFFDDB}0.876 & \cellcolor[HTML]{EAFBE5}12.68 & \cellcolor[HTML]{FFFDDB}0.935 & \cellcolor[HTML]{EAFBE5}0.819 & \cellcolor[HTML]{EAFBE5}132 \\
JLT 64 & 6.5 (4.0x) & 5.9 (4.5x) & 0.2 & \cellcolor[HTML]{F9E0E0}0.846 & \cellcolor[HTML]{EAFBE5}11.04 & \cellcolor[HTML]{FFFDDB}0.909 & \cellcolor[HTML]{EAFBE5}0.842 & \cellcolor[HTML]{F9E0E0}158 \\
JLT 32 & 6.5 (4.0x) & 5.7 (4.5x) & 0.1 & \cellcolor[HTML]{F9E0E0}0.803 & \cellcolor[HTML]{EAFBE5}11.42 & \cellcolor[HTML]{F9E0E0}0.872 & \cellcolor[HTML]{EAFBE5}0.912 & \cellcolor[HTML]{F9E0E0}255 \\ \hline
Mat. 512 & 15.2 (1.7x) & 7.9 (3.3x) & - & \cellcolor[HTML]{EAFBE5}0.892 & \cellcolor[HTML]{EAFBE5}13.89 & \cellcolor[HTML]{EAFBE5}0.947 & \cellcolor[HTML]{EAFBE5}0.808 & \cellcolor[HTML]{EAFBE5}131 \\
Mat. 256 & 8.7 (3.0x) & 6.9 (3.8x) & - & \cellcolor[HTML]{FFFDDB}0.886 & \cellcolor[HTML]{EAFBE5}13.51 & \cellcolor[HTML]{EAFBE5}0.945 & \cellcolor[HTML]{EAFBE5}0.811 & \cellcolor[HTML]{EAFBE5}129 \\
Mat. 128 & 6.2 (4.2x) & 7.0 (3.7x) & - & \cellcolor[HTML]{FFFDDB}0.874 & \cellcolor[HTML]{EAFBE5}12.55 & \cellcolor[HTML]{FFFDDB}0.933 & \cellcolor[HTML]{EAFBE5}0.82 & \cellcolor[HTML]{EAFBE5}130 \\
Mat. 64 & 5.0 (5.2x) & 5.8 (4.5x) & - & \cellcolor[HTML]{F9E0E0}0.843 & \cellcolor[HTML]{EAFBE5}11.07 & \cellcolor[HTML]{FFFDDB}0.906 & \cellcolor[HTML]{EAFBE5}0.843 & \cellcolor[HTML]{F9E0E0}156 \\
Mat. 32 & 4.7 (5.6x) & 5.6 (4.7x) & - & \cellcolor[HTML]{F9E0E0}0.795 & \cellcolor[HTML]{EAFBE5}11.42 & \cellcolor[HTML]{F9E0E0}0.865 & \cellcolor[HTML]{EAFBE5}0.915 & \cellcolor[HTML]{F9E0E0}258 \\ \hline
\rowcolor[HTML]{EFEFEF} 
\multicolumn{9}{c}{\cellcolor[HTML]{EFEFEF}\textbf{MXBAI ($N=769K$, $d=1024$, $k=3508$)}} \\
\rowcolor[HTML]{EFEFEF} 
Raw & 14.3 (1.0x) & 4.8 (3.0x) & - & 0.920 & 8.60 & 0.970 & 101.334 & 97 \\
PCA 512 & 9.3 (1.5x) & 5.0 (2.9x) & 1.4 & \cellcolor[HTML]{EAFBE5}0.921 & \cellcolor[HTML]{EAFBE5}8.60 & \cellcolor[HTML]{EAFBE5}0.970 & \cellcolor[HTML]{EAFBE5}101.233 & \cellcolor[HTML]{EAFBE5}97 \\
PCA 256 & 5.7 (2.5x) & 4.0 (3.5x) & 1.2 & \cellcolor[HTML]{EAFBE5}0.921 & \cellcolor[HTML]{EAFBE5}8.46 & \cellcolor[HTML]{EAFBE5}0.970 & \cellcolor[HTML]{EAFBE5}101.354 & \cellcolor[HTML]{EAFBE5}92 \\
PCA 128 & 4.5 (3.2x) & 3.8 (3.8x) & 1.1 & \cellcolor[HTML]{EAFBE5}0.919 & \cellcolor[HTML]{EAFBE5}8.20 & \cellcolor[HTML]{EAFBE5}0.970 & \cellcolor[HTML]{EAFBE5}101.916 & \cellcolor[HTML]{EAFBE5}88 \\
PCA 64 & 4.2 (3.4x) & 3.8 (3.7x) & 1.1 & \cellcolor[HTML]{EAFBE5}0.912 & \cellcolor[HTML]{EAFBE5}7.93 & \cellcolor[HTML]{EAFBE5}0.967 & \cellcolor[HTML]{FFFDDB}103.450 & \cellcolor[HTML]{EAFBE5}81 \\
PCA 32 & 4.3 (3.3x) & 3.8 (3.7x) & 1.4 & \cellcolor[HTML]{FFFDDB}0.894 & \cellcolor[HTML]{EAFBE5}7.69 & \cellcolor[HTML]{EAFBE5}0.961 & \cellcolor[HTML]{F9E0E0}107.044 & \cellcolor[HTML]{EAFBE5}86 \\ \hline
JLT 512 & 7.6 (1.9x) & 3.9 (3.6x) & 0.5 & \cellcolor[HTML]{EAFBE5}0.919 & \cellcolor[HTML]{EAFBE5}8.51 & \cellcolor[HTML]{EAFBE5}0.969 & \cellcolor[HTML]{EAFBE5}101.421 & \cellcolor[HTML]{EAFBE5}95 \\
JLT 256 & 4.2 (3.4x) & 3.1 (4.5x) & 0.3 & \cellcolor[HTML]{EAFBE5}0.917 & \cellcolor[HTML]{EAFBE5}8.33 & \cellcolor[HTML]{EAFBE5}0.968 & \cellcolor[HTML]{EAFBE5}101.852 & \cellcolor[HTML]{EAFBE5}92 \\
JLT 128 & 3.1 (4.6x) & 2.6 (5.4x) & 0.2 & \cellcolor[HTML]{FFFDDB}0.909 & \cellcolor[HTML]{EAFBE5}7.90 & \cellcolor[HTML]{EAFBE5}0.963 & \cellcolor[HTML]{FFFDDB}102.667 & \cellcolor[HTML]{EAFBE5}90 \\
JLT 64 & 2.6 (5.4x) & 2.6 (5.5x) & 0.2 & \cellcolor[HTML]{FFFDDB}0.889 & \cellcolor[HTML]{EAFBE5}7.11 & \cellcolor[HTML]{FFFDDB}0.949 & \cellcolor[HTML]{FFFDDB}104.984 & \cellcolor[HTML]{EAFBE5}97 \\
JLT 32 & 2.4 (5.8x) & 2.5 (5.6x) & 0.2 & \cellcolor[HTML]{F9E0E0}0.861 & \cellcolor[HTML]{EAFBE5}6.98 & \cellcolor[HTML]{FFFDDB}0.928 & \cellcolor[HTML]{F9E0E0}112.599 & \cellcolor[HTML]{F9E0E0}154 \\ \hline
Mat. 512 & 6.9 (2.1x) & 3.3 (4.3x) & - & \cellcolor[HTML]{EAFBE5}0.919 & \cellcolor[HTML]{EAFBE5}8.50 & \cellcolor[HTML]{EAFBE5}0.969 & \cellcolor[HTML]{EAFBE5}101.507 & \cellcolor[HTML]{EAFBE5}96 \\
Mat. 256 & 3.9 (3.7x) & 2.8 (5.1x) & - & \cellcolor[HTML]{EAFBE5}0.917 & \cellcolor[HTML]{EAFBE5}8.37 & \cellcolor[HTML]{EAFBE5}0.968 & \cellcolor[HTML]{EAFBE5}101.793 & \cellcolor[HTML]{EAFBE5}92 \\
Mat. 128 & 2.9 (4.9x) & 2.6 (5.4x) & - & \cellcolor[HTML]{FFFDDB}0.908 & \cellcolor[HTML]{EAFBE5}7.95 & \cellcolor[HTML]{EAFBE5}0.964 & \cellcolor[HTML]{FFFDDB}102.674 & \cellcolor[HTML]{EAFBE5}89 \\
Mat. 64 & 2.4 (6.0x) & 2.3 (6.2x) & - & \cellcolor[HTML]{FFFDDB}0.892 & \cellcolor[HTML]{EAFBE5}7.19 & \cellcolor[HTML]{FFFDDB}0.952 & \cellcolor[HTML]{FFFDDB}105.051 & \cellcolor[HTML]{EAFBE5}96 \\
Mat. 32 & 2.6 (5.6x) & 2.5 (5.6x) & - & \cellcolor[HTML]{F9E0E0}0.865 & \cellcolor[HTML]{EAFBE5}7.12 & \cellcolor[HTML]{FFFDDB}0.930 & \cellcolor[HTML]{F9E0E0}112.6 & \cellcolor[HTML]{F9E0E0}161 \\ \hline
\end{tabular}
}
\vspace*{-4mm}
\end{table}

The effect of dimensionality reduction on clustering speed is shown in Figure~\ref{fig:eval:dimreduction:speed} and Table~\ref{tab:eval:dimred}. The preprocessing time for PCA is the highest, while remaining negligible relative to the total clustering time. Notably, if the vector search pipeline requires PCA-projected vectors (e.g., MRQ~\cite{happymarriage}), this preprocessing would otherwise occur at index materialization time. On the other hand, Matryoshka representations eliminate preprocessing requirements. It is important to note that speedup is sublinear with respect to dimensionality reduction, a known phenomenon in GEMM routines~\cite{smallgemm}. Remarkably, SuperKMeans delivers modest speedups, ranging from 20\% to 2x. Figure~\ref{fig:eval:dimreduction:speed} shows that SuperKMeans accelerates the clustering of wider vectors to achieve similar speedups to those of thinner vectors without sacrificing quality. Finally, note that SuperKMeans is disabled when $d < 128$, as pruning can be detrimental for performance when approaching this threshold~\cite{superkmeans}.

\vspace*{3mm}
\noindent{\bf Dimensionality Reduction and Vector Search:} Reducing the dimensionality of the raw vectors will affect index materialization, leading to three possible approaches: 1) recompute the full-dimensional centroids using cluster membership (requires access to raw vectors), 2) unproject centroids back to the original dimensionality (not possible when using Matryoshka vectors), and 3) utilize projected centroids for vector search in the reduced space~\cite{wang2026lindormvector}. The results shown in this section use approach \#1.    Table~\ref{tab:eval:dimredsynergy} shows a comparison of all strategies. Both unprojected and projected centroids yield lower clustering quality than recomputed full-dimensional centroids because energy from discarded dimensions is lost. However, the difference between the approaches is minimal at higher levels of preserved variance, allowing the pipeline to avoid accessing the raw vectors entirely after preprocessing. Previous studies have shown the feasibility of using projected vectors and centroids for vector search~\cite{wang2026lindormvector, happymarriage}, although these approaches still require storing the residuals of the PCA projection for re-ranking.

\begin{table}[h]
\renewcommand{\tabcolsep}{1.5pt}
\centering
\caption{Quality of the generated centroids when using three different strategies to materialize the centroids: recomputing centroids at full-d (best), unprojecting the centroids to the original dimensionality, and using the projected centroids. 
}
\vspace*{-4mm}
\label{tab:eval:dimredsynergy}
\resizebox{0.7\columnwidth}{!}{%
\begin{tabular}{lcccc}
\hline
\multicolumn{1}{c}{} &  & \multicolumn{3}{c}{\textbf{Recall@100@1\%}} \\ \cline{3-5} 
\multicolumn{1}{c}{\multirow{-2}{*}{\textbf{Dim.}}} & \multirow{-2}{*}{\textbf{\begin{tabular}[c]{@{}c@{}}Preserved\\ Variance\end{tabular}}} & \textbf{Recomputed} & \textbf{Unprojected} & \textbf{Projected} \\ \hline
\rowcolor[HTML]{EFEFEF} 
\multicolumn{5}{c}{\cellcolor[HTML]{EFEFEF}\textbf{Cohere ($N=10M$, $d=1024$, $k=12649$)}} \\
\rowcolor[HTML]{EFEFEF} 
Raw & 1.00 & 0.884 & 0.884 & 0.884 \\
768 & 0.99 & \cellcolor[HTML]{EAFBE5}0.883 & \cellcolor[HTML]{EAFBE5}0.883 & \cellcolor[HTML]{EAFBE5}0.883 \\
320 & 0.84 & \cellcolor[HTML]{EAFBE5}0.884 & \cellcolor[HTML]{EAFBE5}0.882 & \cellcolor[HTML]{EAFBE5}0.882 \\
256 & 0.78 & \cellcolor[HTML]{EAFBE5}0.884 & \cellcolor[HTML]{EAFBE5}0.880 & \cellcolor[HTML]{EAFBE5}0.880 \\
192 & 0.71 & \cellcolor[HTML]{EAFBE5}0.881 & \cellcolor[HTML]{FFFDDB}0.874 & \cellcolor[HTML]{FFFDDB}0.874 \\
128 & 0.60 & \cellcolor[HTML]{EAFBE5}0.876 & \cellcolor[HTML]{FFFDDB}0.864 & \cellcolor[HTML]{FFFDDB}0.864 \\
64 & 0.43 & \cellcolor[HTML]{FFFDDB}0.865 & \cellcolor[HTML]{F9E0E0}0.826 & \cellcolor[HTML]{F9E0E0}0.826 \\
32 & 0.30 & \cellcolor[HTML]{F9E0E0}0.836 & \cellcolor[HTML]{F9E0E0}0.718 & \cellcolor[HTML]{F9E0E0}0.718 \\ \hline
\end{tabular}
}
\vspace*{-4mm}
\end{table}

\subsection{Mixing Techniques}\label{sec:eval:dimreduction}
\begin{figure}[t!]
\centering
\includegraphics[width=1.0\linewidth]{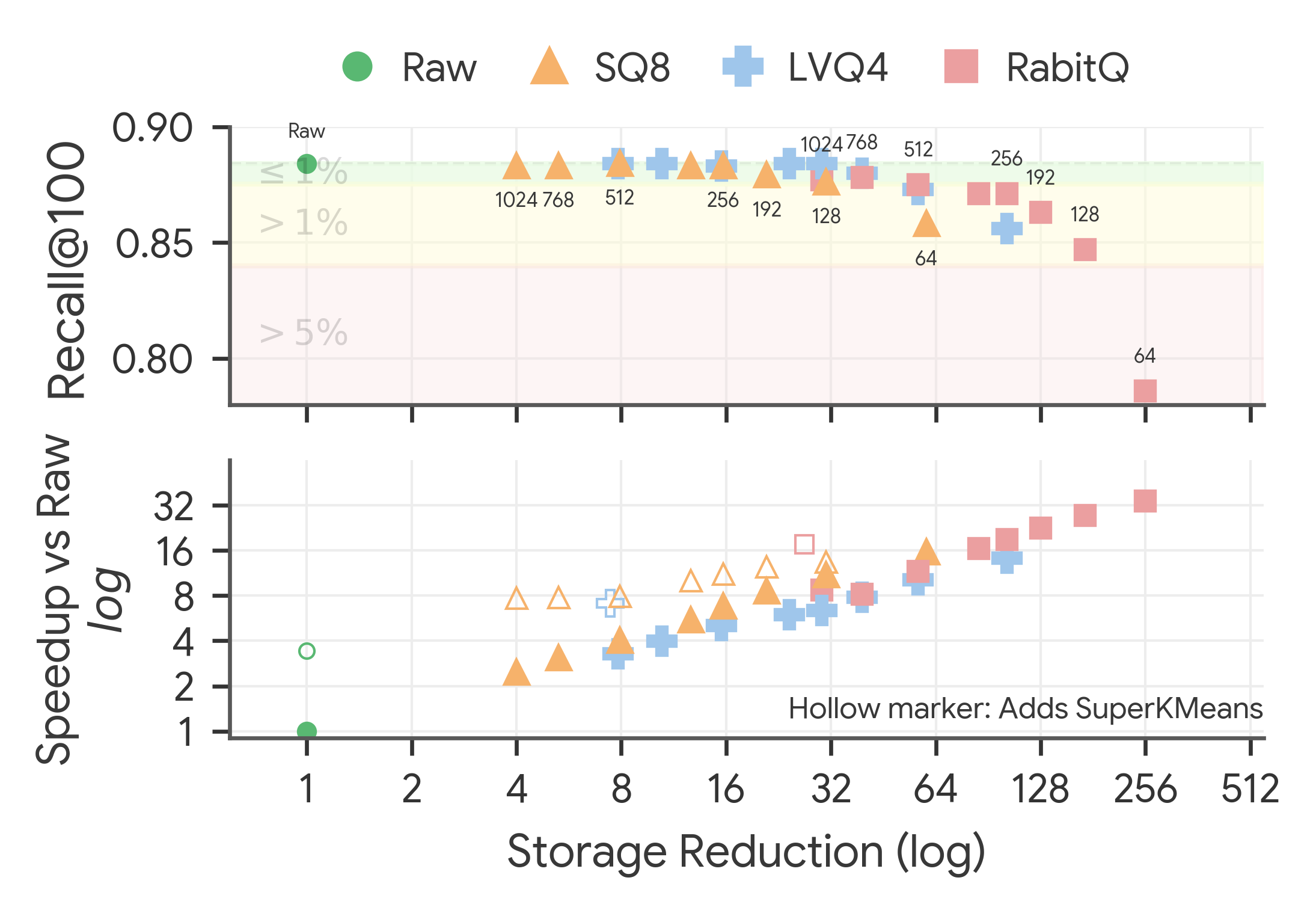}
\vspace*{-9.0mm}
\caption{The effect on clustering quality (top) and speedup (bottom) of combining dimensionality reduction and quantization before clustering. }
\vspace*{-4.0mm}
\label{fig:eval:mix}
\end{figure}

Figure~\ref{fig:eval:mix} shows the result of combining techniques in our largest dataset (Cohere/1024). We present results for SQ8, LVQ4, and RabitQ, combined with PCA as dimensionality-reduction method. If maintaining quality equivalent to clustering raw vectors is essential, LVQ with 80\% of preserved variance is the best choice. However, if quality can deviate by up to 1\% from the optimal value, then RabitQ with 80\% of PCA variance becomes the preferred option. For prioritizing speed while still aiming for near-optimal quality, RabitQ combined with SuperKMeans is the optimal choice. It is worth noting that using RabitQ with PCA and SuperKMeans can degrade performance due to the additional floating-point operations required to compute partial L2 distances for pruning, particularly for thinner vectors. In contrast, SQ8 consistently benefits from SuperKMeans because its distance calculations do not require any additional adjustments.

\subsection{Other Accelerators}


\noindent{\bf Hierarchical k-Means: }
Hierarchical $k$-means~\cite{karypis2000comparison} has recently emerged as an alternative for clustering large collections of vector embeddings~\cite{kmeanselastic, turbopuffer, cuvs, jin2026curator, spfresh, martinico2026efficient, superkmeans} because it reduces the complexity of vanilla $k$-means from $O(N * k * d)$ to $O(N * \sqrt{k} * d)$ by dividing the clustering process into two phases: MESO- and FINE-CLUSTERING. In the MESO-CLUSTERING phase, a coarser clustering is performed, creating only $\sqrt{k}$ clusters. In the subsequent FINE-CLUSTERING phase, $k$-means is applied within each meso-cluster, with the number of clusters set to $\sqrt{n_{i}}$, where $n_{i}$ is the number of points in the i-th meso-cluster. Finally, an optional refinement iteration of vanilla $k$-means can be performed to improve clustering quality further. Table~\ref{tab:eval:hsk} shows the quality of the resulting centroids for vector search tasks on our two largest datasets, Cohere and OpenAI, when using hierarchical $k$-means. The performance gains are remarkable, reaching up to 91x when hierarchical $k$-means is combined with RabitQ. Note that hierarchical $k$-means produces clusters with more uniform sizes in dense regions of the space, as indicated by the lower number of vectors explored. Nonetheless, this improved balance negatively affects recall and requires probing more clusters. 

\pagebreak

\begin{table}[t!]
\renewcommand{\tabcolsep}{1.5pt}
\centering
\caption{Quality of the generated centroids for VSS using hierarchical $k$-means. Hierarchical $k$-means achieves remarkable speedups while producing more balanced clusters.}
\vspace*{-4mm}
\label{tab:eval:hsk}
\resizebox{1.0\columnwidth}{!}{%
\begin{tabular}{lccccccc}
\hline
\textbf{Accelerator} & \textbf{\begin{tabular}[c]{@{}c@{}}Build \\ Time (s)\end{tabular}} & \textbf{\begin{tabular}[c]{@{}c@{}}Recall\\ @100\\ @1\%\end{tabular}} & \textbf{\begin{tabular}[c]{@{}c@{}}Vectors \\ explored \\ @1\% ($\times 10^{3}$)\end{tabular}} & \textbf{\begin{tabular}[c]{@{}c@{}}Recall\\ @100\\ @3\%\end{tabular}} & \textbf{\begin{tabular}[c]{@{}c@{}}Vectors \\ explored \\ @3\% ($\times 10^{3}$)\end{tabular}} & \textbf{\begin{tabular}[c]{@{}c@{}}WCSS \\ ($\times 10^{6}$)\end{tabular}} & \textbf{\begin{tabular}[c]{@{}c@{}}Cluster \\ Size\\ Std.Dev.\end{tabular}} \\ \hline
\rowcolor[HTML]{EFEFEF} 
\multicolumn{8}{c}{\cellcolor[HTML]{EFEFEF}\textbf{Cohere ($N = 10M$, $d = 1024$, $k = 12649$)}} \\
\rowcolor[HTML]{EFEFEF} 
\textbf{Vanilla $k$-means (Raw)} & 640.3 (1.0x) & 0.884 & 105.4 & 0.934 & 313.3 & 5.10 & 354 \\ \hline
\textbf{Hierarchical (Raw)} & 13.6 (47.1x) & \cellcolor[HTML]{FFFDDB}0.854 & \cellcolor[HTML]{EAFBE5}88.1 & \cellcolor[HTML]{FFFDDB}0.914 & \cellcolor[HTML]{EAFBE5}262.8 & \cellcolor[HTML]{FFFDDB}5.26 & \cellcolor[HTML]{EAFBE5}343 \\
\textbf{Hierarchical (SQ8)} & 9.4 (68.1x) & \cellcolor[HTML]{FFFDDB}0.853 & \cellcolor[HTML]{EAFBE5}87.9 & \cellcolor[HTML]{FFFDDB}0.913 & \cellcolor[HTML]{EAFBE5}261.6 & \cellcolor[HTML]{FFFDDB}5.26 & \cellcolor[HTML]{EAFBE5}348 \\
\textbf{Hierarchical (LVQ4)} & 8.2 (78.1x) & \cellcolor[HTML]{FFFDDB}0.853 & \cellcolor[HTML]{EAFBE5}88.4 & \cellcolor[HTML]{FFFDDB}0.913 & \cellcolor[HTML]{EAFBE5}263.8 & \cellcolor[HTML]{FFFDDB}5.26 & \cellcolor[HTML]{FFFDDB}358 \\
\textbf{Hierarchical (RabitQ)} & \textbf{7.0 (91.5x)} & \cellcolor[HTML]{FFFDDB}0.844 & \cellcolor[HTML]{EAFBE5}87.7 & \cellcolor[HTML]{FFFDDB}0.905 & \cellcolor[HTML]{EAFBE5}261.7 & \cellcolor[HTML]{F9E0E0}5.43 & \cellcolor[HTML]{EAFBE5}339 \\ \hline
\rowcolor[HTML]{EFEFEF} 
\multicolumn{8}{c}{\cellcolor[HTML]{EFEFEF}\textbf{OpenAI ($N = 5M$, $d = 1536$, $k = 8944$)}} \\
\rowcolor[HTML]{EFEFEF} 
\textbf{Vanilla $k$-means (Raw)} & 284.8 (1.0x) & 0.952 & 51.9 & 0.982 & 153.7 & 0.99 & 216 \\ \hline
\textbf{Hierarchical (Raw)} & 11.5 (24.8x) & \cellcolor[HTML]{FFFDDB}0.939 & \cellcolor[HTML]{EAFBE5}46.5 & \cellcolor[HTML]{EAFBE5}0.977 & \cellcolor[HTML]{EAFBE5}136.2 & \cellcolor[HTML]{FFFDDB}1.02 & \cellcolor[HTML]{FFFDDB}224 \\
\textbf{Hierarchical (SQ8)} & 8.3 (34.3x) & \cellcolor[HTML]{FFFDDB}0.939 & \cellcolor[HTML]{EAFBE5}46.7 & \cellcolor[HTML]{EAFBE5}0.976 & \cellcolor[HTML]{EAFBE5}137.1 & \cellcolor[HTML]{FFFDDB}1.02 & \cellcolor[HTML]{FFFDDB}220 \\
\textbf{Hierarchical (LVQ4)} & 7.5 (38.0x) & \cellcolor[HTML]{FFFDDB}0.937 & \cellcolor[HTML]{EAFBE5}46.6 & \cellcolor[HTML]{EAFBE5}0.975 & \cellcolor[HTML]{EAFBE5}136.5 & \cellcolor[HTML]{FFFDDB}1.02 & \cellcolor[HTML]{FFFDDB}222 \\
\textbf{Hierarchical (RabitQ)} & \textbf{6.7 (42.5x)} & \cellcolor[HTML]{FFFDDB}0.933 & \cellcolor[HTML]{EAFBE5}46.4 & \cellcolor[HTML]{EAFBE5}0.973 & \cellcolor[HTML]{EAFBE5}136.0 & \cellcolor[HTML]{FFFDDB}1.04 & \cellcolor[HTML]{FFFDDB}220 \\ \hline
\end{tabular}
}
\vspace*{-2mm}
\end{table}

\begin{table}[t!]
\renewcommand{\tabcolsep}{1.5pt}
\centering
\caption{Quality of the generated centroids for VSS tasks when using hierarchical $k$-means and graph-based assignments with HNSW. }
\vspace*{-4mm}
\label{tab:eval:hnsw}
\resizebox{1.0\columnwidth}{!}{%
\begin{tabular}{lcccccccc}
\hline
\textbf{Accelerator} & \textbf{\begin{tabular}[c]{@{}c@{}}Build \\ Time (s)\end{tabular}} & \textbf{\begin{tabular}[c]{@{}c@{}}Build \\ Time (s) \\ w/ SKM\end{tabular}} & \textbf{\begin{tabular}[c]{@{}c@{}}Recall\\ @100\\ @1\%\end{tabular}} & \textbf{\begin{tabular}[c]{@{}c@{}}Vectors \\ explored \\ @1\% ($\times 10^{3}$)\end{tabular}} & \textbf{\begin{tabular}[c]{@{}c@{}}Recall\\ @100\\ @3\%\end{tabular}} & \textbf{\begin{tabular}[c]{@{}c@{}}Vectors \\ explored \\ @3\% ($\times 10^{3}$)\end{tabular}} & \textbf{\begin{tabular}[c]{@{}c@{}}WCSS \\ ($\times 10^{6}$)\end{tabular}} & \textbf{\begin{tabular}[c]{@{}c@{}}Cluster \\ Size\\ Std.Dev.\end{tabular}} \\ \hline
\rowcolor[HTML]{EFEFEF} 
\multicolumn{9}{c}{\cellcolor[HTML]{EFEFEF}\textbf{Cohere ($N = 10M$, $d = 1024$, $k = 12649$)}} \\
\rowcolor[HTML]{EFEFEF} 
\textbf{Raw} & 640.3 (1.0x) & 186.2 (3.4x) & 0.884 & 105.4 & 0.934 & 313.3 & 5.10 & 354 \\ \hline
\textbf{Hierarchical} & 13.6 (47.1x) & 13.4 (47.9x) & \cellcolor[HTML]{FFFDDB}0.854 & \cellcolor[HTML]{EAFBE5}88.1 & \cellcolor[HTML]{EAFBE5}0.914 & \cellcolor[HTML]{EAFBE5}262.8 & \cellcolor[HTML]{FFFDDB}5.26 & \cellcolor[HTML]{EAFBE5}343 \\
\textbf{Hierarchical + Refine} & 28.8 (22.2x) & 26.4 (24.2x) & \cellcolor[HTML]{FFFDDB}0.866 & \cellcolor[HTML]{EAFBE5}91.8 & \cellcolor[HTML]{EAFBE5}0.921 & \cellcolor[HTML]{EAFBE5}274.4 & \cellcolor[HTML]{FFFDDB}5.19 & \cellcolor[HTML]{EAFBE5}331 \\ \hline
\textbf{Graph-based (efs=4)} & \multicolumn{2}{c}{24.3 (26.4x)} & \cellcolor[HTML]{FFFDDB}0.861 & \cellcolor[HTML]{F9E0E0}119.2 & \cellcolor[HTML]{EAFBE5}0.920 & \cellcolor[HTML]{F9E0E0}352.7 & \cellcolor[HTML]{FFFDDB}5.29 & \cellcolor[HTML]{F9E0E0}453 \\
\textbf{Graph-based (efs=8)} & \multicolumn{2}{c}{35.6 (18.0x)} & \cellcolor[HTML]{EAFBE5}0.875 & \cellcolor[HTML]{F9E0E0}111.7 & \cellcolor[HTML]{EAFBE5}0.930 & \cellcolor[HTML]{F9E0E0}331.0 & \cellcolor[HTML]{FFFDDB}5.17 & \cellcolor[HTML]{FFFDDB}371 \\
\textbf{Graph-based (efs=16)} & \multicolumn{2}{c}{59.2 (10.8x)} & \cellcolor[HTML]{EAFBE5}0.881 & \cellcolor[HTML]{FFFDDB}107.5 & \cellcolor[HTML]{EAFBE5}0.932 & \cellcolor[HTML]{FFFDDB}319.1 & \cellcolor[HTML]{EAFBE5}5.13 & \cellcolor[HTML]{FFFDDB}362 \\ \hline
\rowcolor[HTML]{EFEFEF} 
\multicolumn{9}{c}{\cellcolor[HTML]{EFEFEF}\textbf{OpenAI ($N = 5M$, $d = 1536$, $k = 8944$)}} \\
\rowcolor[HTML]{EFEFEF} 
\textbf{Raw} & 284.8 (1.0x) & 89.0 (3.2x) & 0.952 & 51.9 & 0.982 & 153.7 & 0.99 & 216 \\ \hline
\textbf{Hierarchical} & 11.5 (24.8x) & 11.3 (25.3x) & \cellcolor[HTML]{FFFDDB}0.939 & \cellcolor[HTML]{EAFBE5}46.5 & \cellcolor[HTML]{EAFBE5}0.977 & \cellcolor[HTML]{EAFBE5}136.2 & \cellcolor[HTML]{FFFDDB}1.02 & \cellcolor[HTML]{FFFDDB}224 \\
\textbf{Hierarchical + Refine} & 17.8 (16.0x) & 16.7 (17.1x) & \cellcolor[HTML]{EAFBE5}0.945 & \cellcolor[HTML]{EAFBE5}47.6 & \cellcolor[HTML]{EAFBE5}0.980 & \cellcolor[HTML]{EAFBE5}140.5 & \cellcolor[HTML]{FFFDDB}1.01 & \cellcolor[HTML]{FFFDDB}213 \\ \hline
\textbf{Graph-based (efs=4)} & \multicolumn{2}{c}{32.4 (8.8x)} & \cellcolor[HTML]{FFFDDB}0.934 & \cellcolor[HTML]{F9E0E0}58.1 & \cellcolor[HTML]{EAFBE5}0.973 & \cellcolor[HTML]{F9E0E0}175.6 & \cellcolor[HTML]{FFFDDB}1.02 & \cellcolor[HTML]{F9E0E0}254 \\
\textbf{Graph-based (efs=8)} & \multicolumn{2}{c}{51.1 (5.6x)} & \cellcolor[HTML]{EAFBE5}0.946 & \cellcolor[HTML]{FFFDDB}54.2 & \cellcolor[HTML]{EAFBE5}0.979 & \cellcolor[HTML]{FFFDDB}161.6 & \cellcolor[HTML]{EAFBE5}1.00 & \cellcolor[HTML]{FFFDDB}221 \\
\textbf{Graph-based (efs=16)} & \multicolumn{2}{c}{83.8 (3.4x)} & \cellcolor[HTML]{EAFBE5}0.95 & \cellcolor[HTML]{FFFDDB}52.8 & \cellcolor[HTML]{EAFBE5}0.981 & \cellcolor[HTML]{FFFDDB}156.9 & \cellcolor[HTML]{EAFBE5}1.00 & \cellcolor[HTML]{FFFDDB}219 \\ \hline
\end{tabular}
}
\vspace*{-4mm}
\end{table}

\noindent{\bf Graph-based Assignments:} Graph-based indexes for VSS have recently been used to determine assignments during clustering~\cite{seededsearchgraphs, wang2026lindormvector}. This approach builds a graph-based vector index over the centroids (e.g., HNSW~\cite{hnsw}). Consequently, determining assignments becomes a series of top-1 searches on the graph index. Both hierarchical $k$-means and graph-based assignments help clustering scale to larger datasets. Table~\ref{tab:eval:hnsw} compares their performance on Cohere and OpenAI---our two largest datasets. For the graph-based assignment approach, we created an HNSW index using 
\codeword{ef_construction} = 128 and \codeword{M}=16. We use a vanilla implementation of HNSW from the FAISS library~\cite{faisscode}. Additionally, we incorporated an optimization that starts the graph traversal from the centroid assigned in the previous $k$-means iteration~\cite{wang2026lindormvector}. 

Both methods significantly accelerate clustering compared to vanilla $k$-means. However, graph-based assignments produce less balanced clusters, leading to more vectors explored during VSS. Additionally, the graph-based approach introduces the added complexity of tuning the \codeword{ef_search} parameter, which controls search quality. If this value is set too low, it improves speed but compromises clustering quality. Conversely, if set too high, it reduces speed improvements. Note that the graph construction over the centroids takes less than 0.1\% of the total runtime.

\vspace{3mm}
\noindent{\bf Sampling:} Sampling reduces clustering runtime proportionally to the fraction of vectors sampled from the data~\cite{superkmeans, chen2025vectorchord100m}. Sampling around 20--30\% of the data is sufficient to maintain clustering quality and balance~\cite{superkmeans}. Sampling can be used in conjunction with our proposed pipeline. However, note that the final assignment step and index materialization still need to use all vectors. 

\subsection{Scalability}\label{sec:scalability}
\begin{figure}[t!]
\centering
\includegraphics[width=1.0\linewidth]{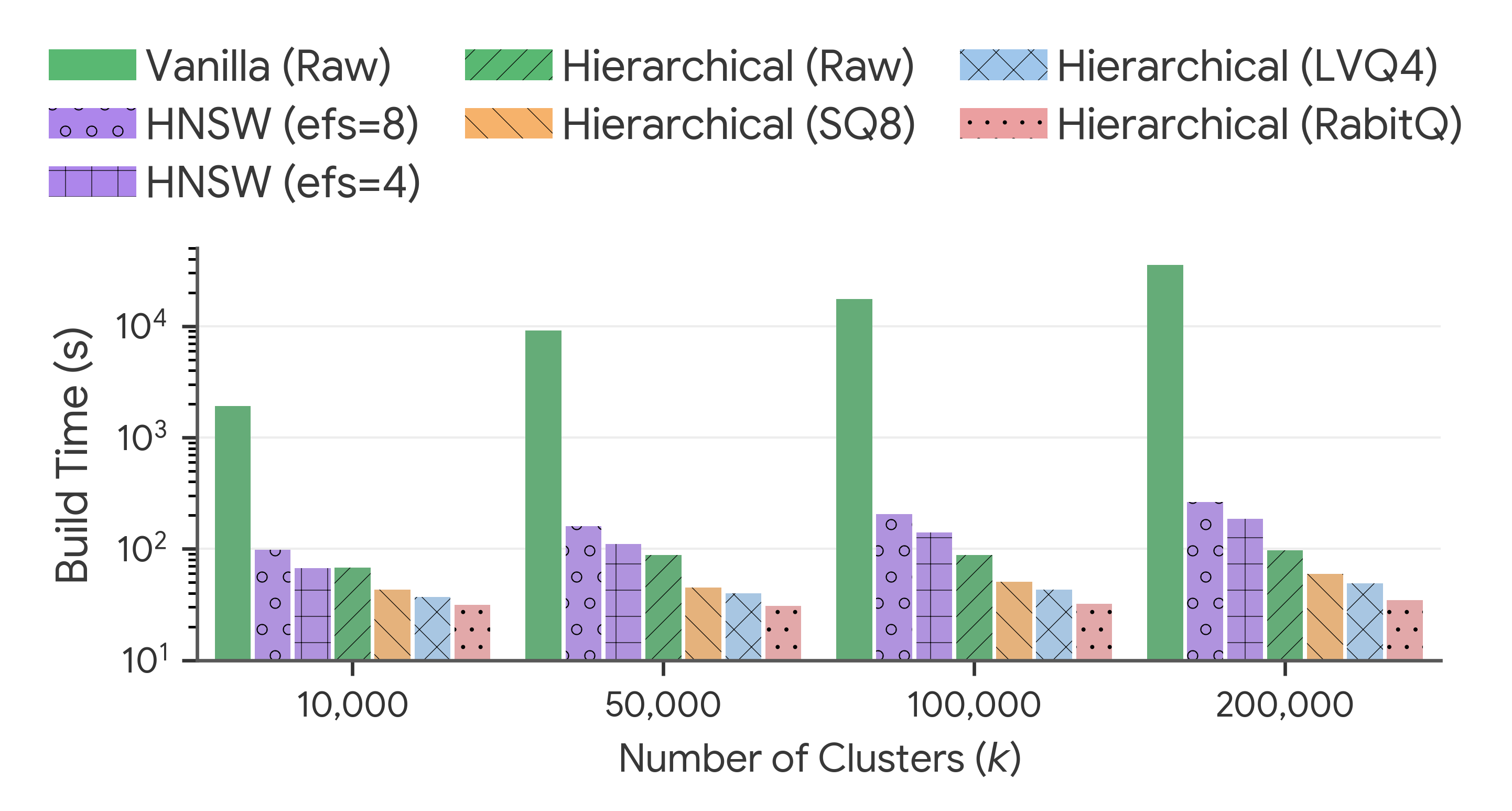}
\vspace*{-8.0mm}
\caption{Clustering runtime scaling with the number of clusters. Both hierarchical $k$-means and graph-based assignments scale sublinearly with $k$. Hierarchical $k$-means with RabitQ provides the lowest clustering time.}
\label{fig:eval:scale}
\end{figure}

The optimal ratio of points per cluster varies depending on the application. Recent studies suggest that a lower points-per-cluster ratio (i.e., a higher $k$) is beneficial for vector search indexes~\cite{wang2026lindormvector}. Consequently, the scalability of our pipeline with respect to $k$ is critical. For this scalability experiment, we scale our Cohere dataset to 50M embeddings~\cite{datasetcohere}, totaling 200GB of data, and run our pipeline while progressively increasing the number of clusters to create ($k$). Additionally, we scale our machine to 512GB of RAM and 64 cores (\codeword{r8a.16xlarge} in AWS). Figure~\ref{fig:eval:scale} shows the results of this experiment. Hierarchical $k$-means with RabitQ vectors is the fastest approach: even when creating 200K clusters, centroid training remains under 1 minute. Notably, both hierarchical $k$-means and graph-based assignments scale sublinearly with respect to the number of clusters. This sublinear scalability is essential for indexing large vector collections, which would otherwise require several hours. The latter, combined with the clustering of quantized vectors, facilitates scalability in both storage requirements and runtime, all while maintaining index quality (as shown in Table~\ref{tab:eval:hsk}).

\section{Discussion}\label{sec:discussion}
Vector indexing through clustering remains the preferred method for large-scale cloud vector systems~\cite{turbopuffer, databricks2026decoupled, scannwhitepaper, xu2025harmony}, where every second of compute counts towards billing. The insights presented in this study show that using full-precision vectors for clustering is excessive: systems can improve the performance of data ingestion pipelines by first applying approximation techniques~\cite{wang2026lindormvector}, wherein method selection can be guided by the algorithms used in the vector search pipeline. For instance, systems using LVQ for vector search~\cite{intelsvs} should apply the encoding before the clustering pipeline.
The same applies to dimensionality reduction techniques, as recent studies propose combining them with quantization~\cite{happymarriage, leanvec, gleanvec, li2025saq, wang2026lindormvector}. Ultimately, our design allows the same vector representations to be used for both vector search and clustering, enabling more streamlined indexing and ingestion in vector systems. The latter opens new research directions toward quantization techniques that can efficiently serve a dual purpose: efficient search \textit{and} indexing. 



\pagebreak

\section{Conclusions and Future Work}\label{sec:conclusion} 

We have introduced a clustering pipeline for indexing vector embeddings in which approximation techniques commonly used for vector search are applied \textit{before} clustering. Our experiments indicate that clustering is highly resilient to approximation techniques and show that even aggressive encodings (RabitQ), combined with dimensionality reduction (PCA) and dimension pruning (SuperKMeans), can reduce storage by up to 60x and substantially accelerate clustering, all while maintaining near-optimal clustering quality. Another striking feat of our study is the adaptation of quantization techniques for clustering pipelines and their integration with dimension pruning. 

In future work, we aim to explore how the availability of specialized hardware units for specific data types (e.g., Intel's AMX) can lead to different trade-offs between speed and storage reduction. Additionally, replicating our study with graph-based indexes (e.g., HNSW, DiskANN) remains a future research opportunity attractive for systems that implement graph-based vector indexes. In the context of the graph-based assignment approach, exploring the performance of different graph indexes or developing novel indexes tailored for this purpose could yield significant benefits. Finally, an evaluation using multi-vector embeddings~\cite{khattab2020colbert} and classic vector datasets that do not stem from AI embedding models (e.g., SIFT, GIST) would broaden the applicability of our pipeline.

\balance
\bibliographystyle{ACM-Reference-Format}
\bibliography{_main}


\appendix



\end{document}